\documentclass[a4paper,11pt]{article}
\pdfoutput=1 

\usepackage{jheppub}

\usepackage{amssymb,amsmath,latexsym,bm,amsfonts}
\usepackage{graphicx}
\usepackage{longtable}
\usepackage{color,xcolor}
\usepackage{indentfirst}
\usepackage{subfigure}

\usepackage[mathscr]{eucal}

\newcommand{\be}{\begin{equation}}
\newcommand{\ee}{\end{equation}}
\newcommand{\bpm}{\begin{pmatrix}}
\newcommand{\epm}{\end{pmatrix}}

\newcommand{\beqn}{\begin{eqnarray}}
\newcommand{\eeqn}{\end{eqnarray}}

\newcommand{\cD}{\mathcal D}
\newcommand{\cG}{\mathcal G}
\newcommand{\cT}{\mathcal T}
\newcommand{\cJ}{\mathcal J}
\newcommand{\cO}{\mathcal O}
\newcommand{\cR}{\mathcal R}

\newcommand{\cS}{\mathcal S}
\newcommand{\cW}{\mathcal W}
\newcommand{\cQ}{\mathcal Q}

\newcommand{\p}{\partial}

\newcommand{\pp}{{++}}
\newcommand{\ppp}{{+++}}
\newcommand{\mm}{{--}}
\newcommand{\mmm}{{---}}

\newcommand{\cN}{{\mathcal N}}
\def\a{\alpha}

\def\d{\delta}
\def\e{\epsilon}

\def\G{\Gamma}

\def\l{\lambda}

\def\q{\theta}

\def\s{\sigma}
\def\t{\tau}

\def\z{\zeta}

\def\F{\Phi}

\def\rd{{\rm d}}

\newcommand{\bea}{\begin{eqnarray}}
\newcommand{\eea}{\end{eqnarray}}
\newcommand{\non}{\nonumber}
\newcommand{\ba}{\begin{array}}
\newcommand{\ea}{\end{array}}

\newcommand{\bsubeq}{\begin{subequations}}
\newcommand{\esubeq}{\end{subequations}}

\def\ri{{\rm i}}

\newcommand{\cDB}{{\bar\cD}}

\newcommand{\pa}{\partial}
\newcommand{\hf}{\frac12}

\newcommand {\cA}{{\cal A}}

\newcommand {\cF}{{\cal F}}
\newcommand {\cH}{{\cal H}}

\newcommand {\cP}{{\cal P}}

\newcommand {\cZ}{{\cal Z}}

\title{Supersymmetric $J\bar{T}$ and $T\bar{J}$ deformations}

\author[a]{Hongliang Jiang,}
\author[a,b]{and Gabriele Tartaglino-Mazzucchelli }

\affiliation[a]{Albert Einstein Center for Fundamental Physics,
Institute for Theoretical Physics,\\
University of Bern,
Sidlerstrasse 5, CH-3012 Bern, Switzerland}

\affiliation[b]{School of Mathematics and Physics, University of Queensland\\
St Lucia, Brisbane, Queensland 4072, Australia}

\emailAdd{jiang@itp.unibe.ch}
 \emailAdd{g.tartaglino-mazzucchelli@uq.edu.au}

\abstract{%
We explore the $J\bar{T}$ and $T\bar{J}$ deformations of 
 two-dimensional field theories possessing 
$\cN=(0,1),(1,1)$ and $(0,2)$ supersymmetry. 
Based on  the  stress-tensor  and flavor current multiplets, we construct various 
  bilinear supersymmetric primary operators 
that induce the $J\bar{T}/T\bar{J}$ deformation in a
manifestly supersymmetric way.
Moreover, their 
supersymmetric descendants
are shown to agree  with the conventional
$J\bar T /T\bar J$ operator on-shell.
We also  present some examples of  $J\bar T /T\bar J$ flows
arising from the supersymmetric deformation of  free theories. 
Finally, we observe that all the deformation operators fit into a general pattern  which generalizes 
the Smirnov-Zamolodchikov type composite operators.}

\begin{document}
\maketitle
\flushbottom


 
\section{Introduction  }

Recently, a new type of deformation of two-dimensional quantum field theories, 
dubbed $T\bar T$ deformation, received a lot of attention.
This arises by deforming a quantum field theory
with a  bilinear composite operator built as the determinant of the stress-energy
tensor~\cite{Zamolodchikov:2004ce,Smirnov:2016lqw,Cavaglia:2016oda}, leading to 
 an \textit{irrelevant} deformation. 
Nevertheless, the $T\bar T$ operator
is free of short-distance divergences
and hence proves to be a well-defined composite local operator \cite{Zamolodchikov:2004ce,Smirnov:2016lqw}.
 Remarkably, the deformation is integrable
and  the spectrum of the deformed theory is related to the undeformed one in a simple 
 way~\cite{Smirnov:2016lqw,Cavaglia:2016oda}. 
 For these reasons, the $T\bar T$ deformation has recently been shown to play an important 
role in many different areas of research and has stirred up excitements in various subjects of high-energy theoretical 
physics.%
\footnote{We do not aim at reviewing here the large bulk of recent research on this subject and we simply refer to 
\cite{Jiang:2019hxb} for a recent, though  not necessarily comprehensive, overview and list of references.}

The $T\bar T$ deformation is the simplest member in a more general family of deformations~\cite{Smirnov:2016lqw}.  Another simple 
member of this family is the $J\bar T/T\bar J $ deformation \cite{Guica:2017lia}, which is constructed out of the stress-energy
 tensor and 
a U(1) current. This deformation explicitly breaks Lorentz invariance, but it still  enjoys several virtues
 of the $T\bar T$ deformation. For 
example, the $J\bar T/T\bar J $ composite operators are also well-defined at the quantum level, 
and they  preserve the solvability enjoyed by the $T\bar T$ deformation. 
Various aspects of the $J\bar T/T\bar J $ deformations have been studied so far, 
including  holography~\cite{Bzowski:2018pcy,Nakayama:2019mvq}, path integral 
formulation~\cite{Aguilera-Damia:2019tpe,Anous:2019osb}, modular invariance~\cite{Aharony:2018ics}, correlation 
functions~\cite{Guica:2019vnb}, and their role in string theory~\cite{Chakraborty:2018vja,Apolo:2018qpq}.   
See also 
\cite{Nakayama:2018ujt,Araujo:2018rho,Giveon:2019fgr,Conti:2019dxg,Chakraborty:2019mdf,Apolo:2019yfj,Hashimoto:2019wct,He:2019vzf,Araujo:2019pae,Frolov:2019xzi}
 for further results.

In this paper, we are going to discuss supersymmetry in the context of  $J\bar T$ and $T\bar J $ deformations. 
The strategy parallels 
with the analysis of supersymmetric $T\bar T$ deformations that was recently discussed 
in a series of papers~\cite{Chang:2018dge,Baggio:2018rpv,Jiang:2019hux,Chang:2019kiu,Ferko:2019oyv}.%
\footnote{See \cite{Coleman:2019dvf} for an alternative geometric,
 though not yet manifestly supersymmetric, method to calculate the $T\bar{T}$ flow of a supersymmetric Lagrangian.} 
There, 
for theories possessing $\cN=(0,1),\,(1,1),\,(0,2)$ and $(2,2)$ supersymmetry,
it was shown how to induce manifestly supersymmetric deformations
in terms of primary operators,
which are constructed out of bilinears of the supercurrent multiplets
(the supersymmetric counterparts  of the stress-energy tensor).
Remarkably, the manifestly supersymmetric deformations 
 prove  to be the same as the ordinary $T\bar{T}$ deformations, up to 
total derivatives and 
 equations of motion.
This result then implies that the ordinary $T\bar{T}$ deformation preserves supersymmetry
and indicates, for example, how to study the $T\bar{T}$-flow of a  Lagrangian  in a manifestly supersymmetric way.

As we are going to show in our paper, all these results can be generalized to the $J\bar T$ and $T\bar J $ case in a similar way. 
As  the starting point, one needs to find the 
supersymmetric counterparts of the stress-energy tensor and U(1) currents. Although the former has been studied extensively,  the 
latter, especially its conservation equation, has scattered results across the literature.
Here we are going to provide a systematic construction of
the flavor current 
multiplets and their conservation equations with $\cN=(0,1),(1,1)$ and $(0,2)$ supersymmetries.
This is done by 
considering the vector multiplets, coupling them to flavor currents, and inspecting their gauge invariances which yield   the 
conservation laws of the flavor current multiplets. See appendix~\ref{DeriveFlavorCurrent} for details.

In the case of $\cN=(0,1),\,(1,1)$ and $(0,2)$ supersymmetry,
by using the stress-tensor multiplets and flavor current multiplets, 
we extend the analysis of \cite{Chang:2018dge,Baggio:2018rpv,Jiang:2019hux}
and construct various supersymmetric primary operators out of their bilinears 
which induce  the  manifestly supersymmetric $J\bar T/T\bar J $ deformations. 
Of particular interest are the cases of chiral supersymmetries,
$\cN=(0,1)$ and $\cN=(0,2)$, where  the supersymmetric extensions of $J\bar{T}$ and $T\bar{J}$ are structurally different.
Like the $T\bar T$ case, 
a fundamental result is that  the  descendants of  the 
$J\bar T/T\bar J $ primary operators 
coincide, on-shell and up to total derivatives, with the conventional $J\bar T/T\bar J $ operators. 
A central aspect of our paper is  to elaborate on these results
and understand in detail the properties of the $J\bar T/T\bar J $ operators.

An interesting observation  arising  from these  analyses is that  
 all the $T\bar T$, $J\bar T$ and $T\bar J$  primary
operators appear to fit into the following general pattern:
     \be\label{ABXY}
\mathcal O =\mathcal A \mathcal B- s\mathcal X\mathcal Y
~, \qquad s=\pm 1~.
 \ee
Here $\mathcal A$, $\mathcal B$, $\mathcal X$ and $\mathcal Y$
are superfields  satisfying  the following constraints
 \be
\mathbb L \mathcal A=  \mathbb R \mathcal  Y, \qquad  \mathbb L \mathcal X=\mathbb R \mathcal B~,
 \ee
 where $\mathbb L,\,\mathbb R$ are  differential operators constructed out 
 of the superspace covariant derivatives.
 These generalize the   Smirnov-Zamolodchikov type composite operators  which corresponds to $\mathbb L=\p_{--},\,\mathbb R=\p_{++},s=1$~\cite{Smirnov:2016lqw}. 
The operator \eqref{ABXY}
 is invariant under improvement transformation with certain assumptions as we show in  appendix~\ref{SZgeneralization}. 
Furthermore, since the  original Smirnov-Zamolodchikov composite   operators were shown to be well-defined at the quantum level~\cite{Smirnov:2016lqw}, we     believe that  the quantum well-definedness  also holds for  our generalized Smirnov-Zamolodchikov type composite   operators~\eqref{ABXY}  with appropriate  $s$.
Indeed, the well-definedness of our  pattern is justified in all the cases considered so far. 
  For $\cN=(0,1),\,(1,1)$ and $(0,2)$ supersymmetric $T\bar{T}$ primary operator, the well-definedness was already elaborated in 
\cite{Chang:2018dge,Baggio:2018rpv,Jiang:2019hux}.  
And for the  $J\bar T/T\bar J$ super-primary operators in this paper,    they will also be shown to be well-defined in appendix~\ref{SZgeneralization}. 
 
 Note that, exactly as in the $T\bar{T}$ case,
the equivalence of the manifestly supersymmetric and the original $J\bar T/T\bar J$  deformations 
ensures that, as far as the analysis of the spectrum goes,
 nothing changes compared to the  results of \cite{Guica:2017lia}; 
 for this reason we avoid to reiterate the analysis of this problem here. 
However, the construction of explicit $J\bar T/T\bar J$-flows for actions and their supersymmetry
is largely sensible to the type of deformation we use.
We will show this feature by constructing some
 $J\bar T/T\bar J $-deformed Lagrangians   explicitly. 
 In particular, we focus on the \emph{chiral} $J\bar T/T\bar J $ 
deformations with $J$ being a chiral  U(1) current, which  was  argued in~\cite{Guica:2017lia} to be the condition of  
solvability.\footnote{However, see  also the very recent paper \cite{Anous:2019osb} 
that solves the spectrum of general $JT_a$  deformations by using a path integral approach.}    
We thus present in our paper several examples of
 chiral $J\bar T$ and $T\bar J $ deformations of free actions with $\cN=(0,1)$  and  $\cN=(0,2)$  
supersymmetry.
 
The paper is organized as follows. In  section~\ref{stresTensorMultiplet}, we set up the notations and  review the stress-tensor 
multiplets with $\cN=(0,1),\, (1,1)$ and $(0,2)$ supersymmetry. 
In section~\ref{flavorCurrentMultiplet}, we present the conservation equations for flavor current multiplets which are derived in 
appendix~\ref{DeriveFlavorCurrent}.
In section~\ref{sec:TJbarsusy}, we construct the primary operators for $J\bar T/T\bar J $ deformations 
and show that their 
descendants coincide with the conventional   $J\bar T/T\bar J $ operator. 
In section~\ref{TJexample}, we discuss some examples of $J\bar T/T\bar J $-deformed   free theories. 
In section~\ref{conclusion}, we conclude and discuss  possible future directions. 
For the reader's convenience, we relegate to two appendices main technical analyses which, however, we believe represent
an important part of our results.
In appendix~\ref{DeriveFlavorCurrent}, we derive  in a systematic way the  conservation equations for the
 flavor current multiplets  with 
$\cN=(0,1),\, (1,1)$ and $(0,2)$ supersymmetry. 
In appendix~\ref{SZgeneralization}, we elaborate on our observation that all the  $T\bar T$, $J\bar T$ and $T\bar J $ deformations 
fit  into the general pattern \eqref{ABXY}  
which goes beyond the Smirnov-Zamolodchikov type of operators. 
In appendix~\ref{SZgeneralization}, we also discuss the well-definedness  of 
the $J\bar T$ and $T\bar J $ primary operators with $\cN=(0,1),\, (1,1)$ and $(0,2)$ supersymmetry.

\section{Stress-tensor multiplets }\label{stresTensorMultiplet}

In this and the next section, we will introduce the stress-tensor multiplets~%
\footnote{They are also commonly called  supercurrent multiplets. But in order to avoid confusion with flavor current multiplets, 
that are  also  supersymmetric current multiplets
and that will be introduced in the next section, 
we will simply call the supercurrent multiplet  as stress-tensor multiplet. } 
and flavor current multiplets with various amount of supersymmetries. These conserved current multiplets are the
building blocks to construct supersymmetric $J\bar T/T\bar J$ operators. 

This section is first devoted to reviewing the stress-tensor multiplet of two-dimensional 
relativistic quantum field theories. After that, since $J\bar T/T\bar J$ deformations break Lorentz invariance, we will also present 
the non-relativistic extensions of the  stress-tensor multiplets.   
This section is also aiming to set up the conventions for the whole paper.

\subsection{$\cN=(0,1)$} \label{01stressTensor}
We begin with  two-dimensional quantum field theories possessing $\cN=(0,1)$ supersymmetry.  
The flat 2D $\cN=(  0,1)$ superspace is parametrized by 
\begin{equation}
\z^M=(\s^{++},\s^{--},\vartheta^+)
~,
\end{equation}
with $\sigma^{\pm\pm}$ being the bosonic light-cone coordinates and $\vartheta^+$ a real Grassmann coordinate.
The spinor covariant derivatives and supercharges are given by
\bea
\cD_+ 
=  \frac{\p}{\p \vartheta^+} - {\rm i}  \vartheta^+ \p_{++}
~,\qquad
\cQ_+ 
= {\rm i}\, \frac{\p}{\p \vartheta^+} -   \vartheta^+ \p_{++}
~,
\eea
and obey the anti-commutation relations
\be
\{ \cD_+ , \cD_+ \} =-2 \ri \p_{++}~,\qquad
\{ \cQ_+ , \cQ_+ \} =-2 \ri \p_{++}~,\qquad
\{\cQ_+, \cD_+\}=0~.
\ee
Given an $\cN=(  0,1)$ superfield $\cF(\z)=\cF(\s,\vartheta^+)$ its supersymmetry transformation is 
\bea\label{01susyTsf}
\d_Q \cF(\z):=-\ri \e_- \cQ_+ \cF(\z)
~,
\label{susy-transf-01_1}
\eea
where $\e_-$ is the constant supersymmetry transformation parameter.
If $F(\s)$ is the operator defined as the $\vartheta=0$ component of the superfield 
$\cF(\zeta)$, $F(\s):=\cF(\s,\vartheta^+)|_{\vartheta=0}$, then 
its supersymmetry transformation is such that
\bea
\d_Q F(\s)=-\ri\e_-\big[Q_+,F(\s)\big\}=-\ri \e_- \cQ_+ \cF(\s,\vartheta^+)\Big|_{\vartheta=0}
=\e_- \cD_+ \cF(\s,\vartheta^+)\Big|_{\vartheta=0}
~.
\label{susy-transf-01_2}
\eea
In our paper we will indicate with $Q_+$ the supersymmetry generator \textit{acting on a component operator}
while $\cQ_+$ is the linear superspace differential operator \textit{acting on superfields}.

For 2D $\cN=(0,1)$ supersymmetric and Lorentz invariant  theories,
the stress-tensor multiplet is described by three superfields $\cT_{----}, \cJ_{+++}$, and $\cJ_-$
satisfying the conservation equations:
\bsubeq
\label{supercurrents-10}
\bea
\cD_+\cT_{----}&=&\ri\p_{--}\cJ_-
~,
\label{supercurrents-10_1}
\\
\p_{--}\cJ_{+++}
&=&-\p_{++}\cJ_-
~.
\label{supercurrents-10_2}
\eea
\esubeq
See \cite{Baggio:2018rpv,Chang:2018dge} for derivations of these  conservation equations 
(either through  the Noether procedure or by requiring the superdiffeomorphism invariance  when coupling to supergravity).
In the superconformal case it holds $ \cJ_-(\zeta)=0 $. 

To describe  the stress-tensor multiplet it is convenient to also define the following two 
descendant superfields
\be
\cT_{++++}:=\cD_+\cJ_{+++}
~, 
~~~~~~
\cT:=\cD_+\cJ_-
~.
\ee
They satisfy
\be
\cD_+\cT_{++++}=-\ri\pa_{++}\cJ_{+++}
~, 
~~~~~~
\cD_+\cT=-\ri\pa_{++}\cJ_-
~,
\ee
and the conservation equations 
\bsubeq
\bea
\p_{++}\cT_{----}&=&-\p_{--}\cT
~,
\label{conservation-cT-10-1}\\
\p_{--}\cT_{++++}&=&-\p_{++}\cT
~.
\label{conservation-cT-10-2}
\eea
\esubeq
The lowest $\vartheta=0$ components of $\cT_{++++}$, $\cT_{----}$ and $\cT$ describe the components of the 
symmetric  stress-energy tensor in light-cone coordinates
\bea
T_{----}(\s)=\cT_{----}(\z)|_{\vartheta=0}
~,~~~
T_{++++}(\s)=\cT_{++++}(\z)|_{\vartheta=0}
~,~~~
\Theta(\s)=\cT(\z)|_{\vartheta=0}
~,~~~~~~
\eea
while the lowest components of $\cJ_{+++}(\z)$ and $\cJ_-(\z)$ define the supersymmetry currents
\be
 J_{+++}(\s)=\cJ_{+++}(\z)|_{\vartheta=0}~, \qquad J_-(\s) =\cJ_-(\z)  |_{\vartheta=0}
 ~.
\ee
In components, the superfields of the stress-tensor multiplet have the following expansion 
\bsubeq
\bea
\cJ_{+++}(\z) &=&J_{+++}(\s)+\vartheta^+ T_{++++}(\s)
~, \\
\cJ_{-}(\z) &=&J_{-}(\s)+\vartheta^+ \Theta(\s)
~, \\
\cT_{----}(\z) &=&T_{----}(\s)+\ri \vartheta^+ \p_{--} J_{-}(\s)
~.
\eea
\esubeq
Due to \eqref{supercurrents-10_1}--\eqref{conservation-cT-10-2},
the operators $T_{\pm\pm\pm\pm}$, $\Theta$, $J_{+++}$ and $J_-$ satisfy the conservation equations
\bsubeq
\begin{align} 
  \{Q_+,J_{+++}\}&=\ri T_{++++}
~,  
& \{Q_+,J_-\}&=\ri\Theta
~,
&~ & 
\\
 {[}Q_+,T_{++++}{]}&=\pa_{++}J_{+++}
~, 
&{[}Q_+,T_{----}{]}&=-\p_{--}J_-
~,  
&{[}Q_+,\Theta{]}&=\pa_{++}J_-
~, 
\\
  \p_{--}J_{+++}
 &=-\p_{++}J_-
~, 
&\p_{++}T_{----}&=-\p_{--}T
~, 
&\p_{--}T_{++++}&=-\p_{++}T
~. 
\end{align}
\esubeq

\subsection{$\cN=(1,1)$}  \label{11stressTensor}

Let us now turn to $\cN=(1,1)$ supersymmetry.
The $\cN=(1,1)$ Minkowski superspace is parametrized by the coordinates
$\zeta^M = (\s^{++},\s^{--},\vartheta^+,\vartheta^-)$.
 The covariant derivatives and supercharges are defined as
\be
\cD_\pm  =  \frac{\p}{\p \vartheta^\pm} - {\rm i}  \vartheta^\pm \p_{\pm\pm}
~,\qquad
\cQ_\pm
 = {\rm i}\, \frac{\p}{\p \vartheta^\pm} -   \vartheta^\pm \p_{\pm\pm}
~,
\label{11DandQ}
\ee
and the anti-commutators read
\bsubeq
\bea
&&
\left\{\cD_\pm,\cD_\pm\right\}  = -2\ri\partial_{\pm\pm}
~,\qquad\quad
\left\{\cQ_\pm,\cQ_\pm\right\}  = -2\ri\partial_{\pm\pm}
~,
\\
&&\left\{\cD_+,\cD_-\right\}  = \left\{\cD_\pm,\cQ_\pm\right\}=\left\{\cQ_+,\cQ_-\right\}  = 0
~.
\eea
\esubeq
The definition of the $\cN=(1,1)$ supersymmetry transformations of an $\cN=(1,1)$ superfield and its lowest component, 
and accordingly the definition of the generators $Q_\pm$ acting on component operators,
is a straightforward extension of the $\cN=(0,1)$ case, see eqs.~\eqref{susy-transf-01_1}--\eqref{susy-transf-01_2},
where supersymmetry transformations are parametrized by $\e_{\pm}$ in the $\cN=(1,1)$ case.

Field theories that are $\cN=(1,1)$ supersymmetric and Lorentz invariant  
possess two pairs of superfields, $(\cJ_{+++}(\z),\cJ_{-}(\z))$ and 
$(\cJ_{---}(\z),\cJ_{+}(\z))$, which describe the stress-tensor multiplet.
The conservation equations are encoded in the following equations (see \cite{Baggio:2018rpv,Chang:2018dge} for 
recent derivations)\footnote{In the supergravity approach~\cite{Baggio:2018rpv}, it holds
$\cJ_\pm(\zeta) =\mp \ri \cD_\pm \cJ (\zeta)$.
}
\begin{equation}
\label{supercurrents-11}
\cD_+\cJ_{---}= \cD_-\cJ_-
~,~~~~~~
\cD_-\cJ_{+++}=\cD_+\cJ_+
~,~~~~~~
\cD_+\cJ_-=\cD_-\cJ_+:=\cT
~.
\end{equation}
We define the following descendant superfields 
\be
\cT_{\pm\pm\pm\pm}:=\cD_\pm\cJ_{\pm\pm\pm}~,\qquad
\cZ_{\pm\pm}:=\cD_\pm\cJ_{\pm}
~.
\ee
Such definitions, together with eq.~\eqref{supercurrents-11}, imply
\bsubeq
\begin{align}
 \cD_\pm\cT_{\pm\pm\pm\pm}&=-\ri\p_{\pm\pm}\cJ_{\pm\pm\pm}
~, 
&\cD_\mp\cT_{\pm\pm\pm\pm}&=\ri \p_{\pm\pm}\cJ_\pm
~, 
&\cD_{\pm}\cT&=-\ri\pa_{\pm\pm}\cJ_\mp
~,
\\
 \cD_\pm\cZ_{\pm\pm}&=-\ri\p_{\pm\pm}\cJ_{\pm}
~, 
&\cD_\mp\cZ_{\pm\pm}&=\ri \p_{\pm\pm}\cJ_\mp
~,
\\
 \partial_{\mp\mp}\cJ_{\pm\pm\pm}  &= - \partial_{\pm\pm}\cJ_{\mp}
~,\ 
&\p_{\pm\pm}\cT_{\mp\mp\mp\mp} &= -\p_{\mp\mp}\cT
~,
&\pa_{--}\cZ_{++} &=-\pa_{++}\cZ_{--}
~.
\end{align}
\esubeq

It is clear that $\cJ_{\pm\pm\pm}$ and $\cJ_\pm$
belong to a stress-tensor multiplet where $\cJ_\pm$ play the role of the supertrace. 
In fact, if the matter system is superconformal then it holds $ \cJ_\pm=0$.
The $\vartheta^\pm=0$ components of $\cJ_{\pm\pm\pm}(\z)$ and $\cJ_\pm(\z)$ define the 
supersymmetry currents $J_{\pm\pm\pm}(\s):=\cJ_{\pm\pm\pm}(\zeta)|_{\vartheta^\pm=0}$ and
 $J_\pm(\s):=\cJ_\pm(\zeta)|_{\vartheta^\pm=0} $, respectively.
The lowest component of $\cZ_{\pm\pm}$, $Z_{\pm\pm}(\s):=\cZ_{\pm\pm}(\z)|_{\vartheta^\pm=0}$, define a central charge current.
The components of the symmetric  stress-energy tensor in light-cone coordinates can be defined as
\bsubeq
\bea
T_{\pm\pm\pm\pm}(\s)
&:=&
\cT_{\pm\pm\pm\pm}|_{\vartheta^{\pm}=0}  
=\cD_\pm \cJ_{\pm\pm\pm}|_{\vartheta^{\pm}=0}  
~,\\
 \Theta(\s)
 &:=&
 \cT|_{\vartheta^{\pm}=0}= \cD_+ \cJ_{-}|_{\vartheta^{\pm}=0}=\cD_- \cJ_{+}|_{\vartheta^{\pm}=0}
~.
\eea
\esubeq
The expansion in components of $\cJ_{\pm\pm\pm}$ and $\cJ_\pm$ read
\bsubeq
\bea
\cJ_{\pm\pm\pm}(\z)&=&
J_{\pm\pm\pm}(\s)
+\vartheta^\pm T_{\pm\pm\pm\pm}(\s)
+\vartheta^\mp Z_{\pm\pm}(\s)
\pm\ri\vartheta^+\vartheta^-
\,\pa_{\pm\pm}J_\pm(\s)
~,
\\
\cJ_{\pm}(\z)
&=&
J_{\pm}(\s)
+\vartheta^\pm Z_{\pm\pm}(\s)
+\vartheta^\mp \Theta(\s)
\pm\ri\vartheta^+\vartheta^-
\pa_{\pm\pm}J_\mp(\s)
~.
\eea
\esubeq
It is straightforward to prove that 
the operators $T_{\pm\pm\pm\pm}$, $\Theta$, $Z_{\pm\pm}$, $J_{\pm\pm\pm}$ and $J_\pm$ 
satisfy the conservation equations
\bsubeq
\begin{align}
 \{Q_\pm,J_{\pm\pm\pm}\}&=\ri T_{\pm\pm\pm\pm}
~, \qquad
 \{Q_\pm,J_\mp\} =\ri\Theta
~, 
&  {[}Q_\pm,T_{\pm\pm\pm\pm}{]}&=\pa_{\pm\pm}J_{\pm\pm\pm}
~,  
\\ 
\{Q_\pm,J_\pm\}   &=    \{Q_\mp,J_{\pm\pm\pm}\}=\ri Z_{\pm\pm}
~,  
&{[}Q_\mp,T_{\pm\pm\pm\pm}{]}&=-\p_{\pm\pm}J_\pm
~, 
\\
{[}Q_\pm,\Theta{]}&=-{[}Q_\mp Z_{\pm\pm}{]}=\pa_{\pm\pm}J_\mp
~, 
&{[}Q_\pm,Z_{\pm\pm}{]}&=\pa_{\pm\pm}J_{\pm}
~,
\\
\p_{\mp\mp}J_{\pm\pm\pm}
&=-\p_{\pm\pm}J_\mp
~, \quad
 \p_{\mp\mp}T_{\pm\pm\pm\pm} =-\p_{\pm\pm}T
~,  
&\p_{--}Z_{++}&=-\p_{++}Z_{--}
~. 
\end{align}
\esubeq

\subsection{$\cN=(0,2)$}  \label{02stressTensor}

Finally, we discuss the  case of $\cN=(0,2)$ supersymmetry. 
Its Minkowski superspace is parametrized by 
$\zeta^M = (\s^{++},\s^{--},\vartheta^+,\bar\vartheta^+)$
with $\vartheta^+$ now a complex Grassmann coordinate and $\bar\vartheta^+= \overline{(\vartheta^+)}$. 

 The covariant derivatives and supercharges are defined as (for later convenience in deriving the flavor current multiplet, 
 we follow the notation in \cite{Brooks:1986gd})
\bsubeq\label{02DandQ}
\bea
\cD_+&=&\frac{\p}{\p \vartheta^+} +\ri\bar \vartheta^+ \p_{++}~,\qquad  \bar\cD_+=\frac{\p}{\p \bar \vartheta^+} 
+\ri  \vartheta^+ \p_{++}
~,
\\
\cQ_+&=&\ri\frac{\p}{\p \vartheta^+} +\bar \vartheta^+ \p_{++},\qquad  \bar\cQ_+=\ri\frac{\p}{\p \bar \vartheta^+} + \vartheta^+ \p_{++}
~,
\eea
\esubeq
satisfying the following (anti-)commutation relations
\be
\cD_+^2=\bar \cD_+^2=0
~, \qquad  \{\cD_+ , \bar \cD_+ \}=2\ri \p_{++}
~, \qquad   [\cD_+ , \p_{\pm \pm} ]=[\bar\cD_+ , \p_{\pm \pm} ]=0
~,
\ee
with equivalent relations satisfied by $\cQ_+$, $\bar{\cQ}_+$ and $\pa_{\pm\pm}$.

For a Lorentz and $\cN=(0,2) $ supersymmetric theory
the general  stress-tensor  multiplet, or supercurrent ``$\cS$-multiplet'', 
was studied in \cite{Dumitrescu:2011iu}. In terms of our notation, the $\cS$-multiplet is determined by the following 
differential constraints~%
\footnote{Note that for simplicity we set to zero the $\cS$-multiplet space-filling brane charge appearing in the constraint 
$\bar\cD_+  \cW_- =C$
since it is linked to supersymmetry breaking \cite{Hughes:1986dn,Dumitrescu:2011iu}.}
\bsubeq
\label{02-Smultiplet}
\bea
\p_\mm\cS_\pp &=&  \cD_+ \cW_- +\bar\cD_+ \bar \cW_- 
~,
\label{02-Smultiplet-1} \\
 \bar \cD_+ \cT_{----} &=& \p_{--}  \cW_-
 ~, 
 \label{02-Smultiplet-2}\\
\cD_+ \cT_{----} & =& - \p_{--} \bar\cW_- 
~,
\label{02-Smultiplet-3}\\
\bar\cD_+  \cW_- &=& \cD_+\bar \cW_-=0
~.
\label{02-Smultiplet-4}
\eea
\esubeq
In our notation, the component   expansions   of the superfields $\cS_\pp$, $\cT_{----}$ and $\cW_-$
solving the previous constraints are given by
\bsubeq
\label{02-Smultiplet-components}
\beqn
 \cT_{----}(\z) 
 &=&
 T_{----}(\s) 
 +\frac12 \vartheta^+ \p_{--} S_{+--}(\s)
 -\frac12 \bar \vartheta^+ \p_{--}\bar S_{+--}(\s)
\non\\
&&
 -\frac{1}{2} \vartheta^+\bar\vartheta^+  \p_\mm^2 j_\pp(\s) 
 ~,\\
  \cS_\pp(\z)
  &=&
  j_\pp(\s) 
  +\ri\vartheta^+ S_{+++}(\s) 
  + \ri\bar \vartheta^+ \bar S_{+++}(\s) 
  + 2\vartheta^+\bar \vartheta^+ T_{++++}(\s)
  ~, \\
  \cW_-(\z)
  &=&
  -\frac12 \bar S_{+--}(\s) 
  -\ri\vartheta^+ \left[\, \Theta(\s) +\frac{\ri}{2} \p_\mm j_{++}(\s)\right]
    -\frac{\ri}{2} \vartheta^+\bar\vartheta^+ \p_{++} \bar S_{+--}(\s)
  ~.
  \qquad \qquad 
\eeqn
\esubeq
The operators $T_{\pm\pm\pm\pm}$ and $\Theta$ are the light-cone components of the symmetric stress-tensor
while $S_{+\pm\pm}$ and its conjugate $\bar{S}_{+\pm\pm}$ are the $\cN=(0,2)$ supersymmetry currents.
They satisfy the conservation equation
\be
 \p_{\mp\mp} T_{\pm\pm\pm\pm}= -\p_{\pm\pm} \Theta
~,\qquad
 \p_{++} S_{+--} = - \p_{--} S_{+++}
 ~.
\ee
Note that, by using \eqref{02-Smultiplet}--\eqref{02-Smultiplet-components}, as for the $\cN=(0,1)$ and $\cN=(1,1)$ cases,
it is straightforward to compute the action of the 
$Q_+$ and $\bar{Q}_+$ generators on the component fields of the $\cS$-multiplet.\footnote{Here for a superfield $\cF(\zeta)$ with 
lowest component $F(\sigma)=\cF(\zeta)|_{\vartheta=0}$, the 
supersymmetry transformations 
act on $F(\s)$ as  $\, \delta_Q F(\sigma) 
=-\ri \big[\epsilon_-Q_+ +\bar\epsilon_- \bar Q_+, F(\sigma)\big\}=\d_Q \cF |_{\vartheta=0 } \, $
 with $\, \d_Q \cF(\z):= -\ri  \big( \e_- \cQ_+ +\bar\e_- \bar\cQ_+ \big) \cF(\z)$.
}

We also define the descendant superfields
\be
\cT_{++++}:=\frac14[ \bar \cD_+, \cD_+ ]\cS_{++}
~, \qquad
\cT:=\frac{\ri}{2} \Big( \cD_+\cW_- -\bar \cD_+\bar \cW_- \Big)
~,
\label{descendants-S-multiplet}
\ee
whose lowest components are $T_{++++}$ and $\Theta$.

One can then check that it holds: 
\be
\cD_+ \Big(\p_{--} \cS_{++} -2\ri \cT \Big) =0~, \qquad
\bar\cD_+ \Big(\p_{--} \cS_{++} +2\ri \cT \Big) =0
~,
\ee
and
\be
[\bar\cD_+, \cD_+]\cT=\p_{++}\p_{--} \cS_{++}
~.
\ee

The $\cS$-multiplet is in general reducible \cite{Dumitrescu:2011iu}.
For instance, for $\cN=(0,2)$ supersymmetric theories admitting a conserved $U(1)_R$ $R$-symmetry, 
the $\cS$-multiplet can be improved to the so-called $\cR$-multiplet.
In this case, the superfield currents $\cW_-(\z)$ and $\bar{\cW}_-(\z)$ are the descendants of a real superfield $\cR_{--}(\z)$
 \be
 \cW_-=\frac{\ri}{2} \bar \cD_+ \cR_{--}
 ~, \qquad 
\bar  \cW_-=  \frac{\ri}{2} \cD_+ \cR_{--}
~.
\label{calW=DR}
 \ee
Then, once we redefine  $\cS_{++}(\z) \equiv\cR_{++}(\z)$ for the $\cR$-multiplet, the conservation equations
\eqref{02-Smultiplet} turn into%
\bsubeq\label{02Rmulti}
\bea
\p_{--} \cR_{++}
&=&
-\p_{++}\cR_{--}
~,
\label{02Rmulti-1}
\\
\bar \cD_+\Big(\cT_{----}-\frac{\ri}{2} \p_{--} \cR_{--}\Big)
&=& 
 \cD_+\Big(\cT_{----}+\frac{\ri}{2} \p_{--} \cR_{--}\Big)=0
 ~.
 \label{02Rmulti-2}
\eea
\esubeq
The conserved vector $R$-symmetry current is then given by the component operators 
$j_{++}(\s)=\cR_{++}(\z)|_{\vartheta=0}$ and
$j_{--}(\s)=\cR_{--}(\z)|_{\vartheta=0}$ such that
\be
\p_{--} j_{++}
=
-\p_{++}j_{--}
~.
\ee
See \cite{Dumitrescu:2011iu} for more detail and \cite{Jiang:2019hux} for a recent derivation by using $\cN=(0,2)$ supergravity.
For the $\cR$-multiplet, note that following useful relation, which derive from \eqref{02Rmulti},
also hold
\be
\cT=\frac14 [\bar \cD_+ , \cD_+ ]\cR_{--}~, \qquad
[\bar \cD_+ , \cD_+ ]\cT_{----}=\p_{++}\p_{--} \cR_{--}
~.
\label{D+D+T----02}
\ee
 
To conclude, note that if the field theory is $\cN=(0,2)$ superconformal 
then it holds $\cW_-=\bar{\cW}_-=0$ and the $\cS$-multiplet is accordingly further simplified.

\subsection{Caveat on the  non-Lorentz-invariant case}

In the previous subsections, we have reviewed the stress-tensor multiplets 
of relativistic quantum field theories possessing various types of supersymmetries.  
However, since $T\bar J$ and $J\bar T$  deformations break Lorentz invariance, 
the deformed theory is not Lorentz invariant any longer. For this reason, in this subsection we are going 
to extend the description of the stress-tensor multiplets to non-Lorentz-invariant supersymmetric field theories.

Given a supersymmetric theory, since translational and supersymmetry invariance are always preserved, 
according to the Noether theorem, the  stress-energy tensor   and supercharges are always well-defined and conserved. 
However, if Lorentz invariance is missing,
the  stress-energy tensor is no longer symmetric --- in light-cone coordinates the two off-diagonal components of 
the stress-energy tensor
\be 
\Theta(\s)=T_{++--}(\s)
~, \qquad \tilde{\Theta}(\s):= T_{--++}(\s)
~,
\label{Theta-ThetaTilde}
\ee
are independent $\Theta(\s)\ne\tilde{\Theta}(\s)$. Translation invariance implies the conservation
 equations for the pairs of currents $(T_{++++},\,\Theta)$ and $(T_{----},\,\tilde\Theta)$
 separately
 \be
 \pa_{--} T_{++++}=-\pa_{++}\Theta
 ~,\qquad
 \pa_{++} T_{----}=-\pa_{--}\tilde{\Theta}
 ~.
\ee
If $\Theta$  or $\tilde\Theta$ are zero then the field theory possesses a chiral $SL(2,\mathbb{R})$ symmetry, 
which enhances to a chiral Virasoro algebra.
Despite these differences, it is straightforward to extend the analysis of the supersymmetric stress-tensor multiplets.
In fact, as we are now going to describe, up to appropriately distinguishing $T_{++--}$ and $T_{--++}$, 
we can effectively use the results of the relativistic field theories.

 \subsubsection*{Non-Lorentz-invariant $ \cN=(0,1)$}

When there are Lorentz anomalies, the supercurrents were discussed, for example, in \cite{Smailagic:1992pj}. 
The conservation equations corresponding to different symmetries are given by: 

\begin{enumerate}

\item Translation invariance:
\bsubeq\beqn\label{nonLorenCon1}
\cD_+ \cT_{----} &=& \ri\p_{--} \tilde \cJ_-
~, \\
\p_{--} \cJ_{+++} &=&-\p_{++} \cJ_- 
~. \label{nonLorenCon2}
\eeqn
\esubeq

\item Dilatation invariance:
\be
\tilde \cJ_- +\cJ_-=0
~.
\ee

\item Lorentz invariance:
\be
\tilde \cJ_- -\cJ_-=0
~.
\ee
\end{enumerate}

Hence, if we require  Lorentz invariance, we obtain exactly the  conservation equations we discussed before for  
relativistic quantum field theory.

To consider the deformation of non-Lorentz invariant field theory, we can only use the first two equations \eqref{nonLorenCon1} 
and \eqref{nonLorenCon2}
and remain with an independent set of superfield currents given by
$(\cT_{----}(\z),\,\cJ_{+++}(\z),\,\cJ_-(\z),\,\tilde{\cJ}_-(\z))$. 
In this case, the 
 stress-energy tensor is not symmetric:
\be
\Theta \equiv \cT|_{\vartheta=0}\neq\tilde \cT|_{\vartheta=0} \equiv\tilde\Theta
~,
\ee
with $
\cT= \cD_+ \cJ_-, \,  \tilde \cT= \cD_+ \tilde \cJ_-
$.

 \subsubsection*{Non-Lorentz-invariant $ \cN=(1,1)$}

Similarly, in the $\cN=(1,1)$ case, the conservation equations corresponding to different symmetries are given by: 

\begin{enumerate}

\item Translation invariance:
\bsubeq\beqn\label{N11con1}
\cD_+ \cJ_{ ---} &=& -\p_{--} \tilde \cJ 
~, \\
\cD_- \cJ_{+++} &=& -\p_{++ }   \cJ 
~. \label{N11con2}
\eeqn
\esubeq

\item Dilatation invariance:
\be
\tilde \cJ  +\cJ =0
~.
\ee

\item Lorentz invariance:
\be
\tilde \cJ  -\cJ =0
~.
\ee
\end{enumerate} 
For Lorentz invariant theory, we thus have $\tilde \cJ  -\cJ =0$, and \eqref{N11con1}, \eqref{N11con2}  reduce  to  
\eqref{supercurrents-11} with $\cJ_\pm =\mp \ri \cD_\pm \cJ  $. While in our non-Lorentz invariant field theories, we should use 
\eqref{N11con1} and \eqref{N11con2} while
keeping ($\tilde \cJ_-(\z) \ne\cJ_-(\z)$) $\tilde \cJ \ne\cJ$.

 \subsubsection*{Non-Lorentz-invariant $ \cN=(0,2)$}

 To accommodate the fact that for non-Lorentz invariant theories 
 the stress-energy tensor is not necessarily symmetric, it turns out that
the $ (0,2)$ $\cS$-multiplet  constraints \eqref{02-Smultiplet} should be modified as follows 
\bsubeq\beqn
\p_\mm\cS_\pp &=&  \cD_+  \cW_- +\bar\cD_+ \bar  {  \cW}_-
~,  \\
 \bar \cD_+ \cT_{----} &=& \p_{--}  \widetilde \cW_-
 ~, \\
\cD_+ \cT_{----} & =& - \p_{--} \bar{ \widetilde\cW}_-
~, \\
\bar\cD_+  \cW_- &=& \cD_+\bar \cW_-=\bar\cD_+  \widetilde\cW_-  =  \cD_+\bar  { \widetilde\cW}_-=0
~,
\eeqn\esubeq
with 
\bsubeq
\beqn
 \cW_-&=& -\ri\vartheta^+ \Big( \Theta+\frac{\ri}{2} \p_\mm j_{++}\Big)+\cdots~,\qquad
  \bar \cW_-=   \ri \bar\vartheta^+ \Big( \Theta -\frac{\ri}{2} \p_\mm j_{++}\Big)+\cdots
  ~, \qquad
\\  \widetilde\cW_-&=& -\ri\vartheta^+ \Big( \tilde{\Theta} +\frac{\ri}{2} \p_\mm j_{++}\Big)+\cdots,\qquad
 \bar  { \widetilde\cW}_-=   \ri \bar\vartheta^+ \Big( \tilde{\Theta} -\frac{\ri}{2} \p_\mm j_{++}\Big)+\cdots
 ~.\qquad
\eeqn
\esubeq
For the multiplet, 
the dots above, and  all other superfields, are the same as the ones in \eqref{02-Smultiplet-components}. 
The $(0,2)$ $\cR$-multiplet for a non-Lorentz invariant theory can similarly be derived in a straightforward way 
from the Lorentz invariant case and we leave to the reader the details for its derivation.

It is crucial to emphasize that these modifications for non-Lorentz-invariant theories are actually not needed for our 
$T\bar J /J\bar T$-deformations. In fact, as we will see later in our analysis,
these composite operators always involve  \emph{only  one} off-diagonal component, either $\Theta$ or $\tilde \Theta$.%
\footnote{Thought it will not play a role in our discussion,
note that the difference between $\Theta$ and $\tilde \Theta$
might instead be relevant for other types of composite operators, for example the $T\bar T$-operator in  
 non-Lorentz invariant field theories  discussed in \cite{Cardy:2018jho}. }

Since this difference proves to be irrelevant 
for our analysis, we will ``pretend" to be working in relativistic theories, 
namely we will just use the normal conservation equations described in the last subsections without 
using the tildes when we refer to $\tilde \cJ_-, \tilde \cJ$ or $ \widetilde\cW_-$.


\section{Flavour current multiplets} \label{flavorCurrentMultiplet}

To construct the supersymmetric $T\bar J/J\bar T$ primary operators, we also need to  derive the supercurrent multiplet for a 
gauge/flavour symmetry. For simplicity, we will restrict to the Abelian case with U(1) 
symmetry. 

The flavor current multiplet can be found in a standard fashion as follows. 
For a given amount of supersymmetry, we first need to  find the gauge multiplet as well as their gauge transformations 
rules, then we couple the gauge multiplet to the corresponding flavor current multiplet. To linearized order, the gauge invariance of 
gauge-current couplings gives rise to the conservation equations of the flavor current multiplets. 
 We defer the details of the derivations 
to appendix~\ref{DeriveFlavorCurrent} and  here present only the final results for the conservation equations.

\subsection{$\cN=(0,1)$} \label{01flavor}

For quantum field theories with $\cN=(0,1)$  supersymmetry, as derived in appendix~\ref{DeriveFlavorCurrent01},  the  flavor current 
multiplet of an Abelian 
symmetry consists of two superfields $\cG_{--}(\z)$ and $\cG_+(\z)$ satisfying the following constraint:
\bea\label{flavorcurrent01}
\cD_+\cG_{--}=\ri\p_{--}\cG_+
~.
\eea
If we  define 
\bea
\cG_{++}:=\cD_+\cG_+
~,
\eea
then we have   
\bea
\cD_+\cG_{++}=-\ri\p_{++}\cG_+
~,
\eea
and 
\bea
\p_{++}\cG_{--}=-\p_{--}\cG_{++}
~.
\eea

In components, the current multiplet is given by
\be
\cG_{--}(\z)=G_{--}(\s)+ \ri \vartheta^+ \p_{--} g_+(\s)
~, \qquad \cG_+(\z)=g_+(\s) +\vartheta^+ G_{++}(\s)
~,
\ee
whose components $G_{\pm\pm}(\s)$ satisfy the conservation equation for a vector current
\be
\p_{++} G_{--}+\p_{--} G_{++}=0
~.
\ee

\subsection{$\cN=(1,1)$}\label{11flavor}

As shown in appendix~\ref{DeriveFlavorCurrent11}, the  flavor current multiplet   consists of
two superfields $\cG_{-}(\z)$ and $\cG_+(\z)$ satisfying the following constraint: 
\be\label{flavorCurrent11}
   \cD_+ \cG_-  -   \cD_- \cG_+=0
   ~.
 \ee
Acting with $\cD_+\cD_-$ on  both sides of the previous equation gives
\be
\p_{++} \cG_{--}+\p_{--}\cG_{++}=0
~,
\ee
where we have defined the following descendant superfields
 \be
  \cG_{++}=  \cD_+ \cG_+
  ~, \qquad     
  \cG_{--}=  \cD_- \cG_-
  ~.
 \ee
The flavor current multiplet can be expressed in terms of component fields as: 
\bsubeq
\beqn
\cG_+(\z) &=&  g_+(\s) +\vartheta^+ G_{++}(\s) +\vartheta^- p(\s) +\ri \vartheta^+\vartheta^- \p_{++} g_-(\s)   
~, \\
\cG_-(\z) &=&  g_-(\s) +\vartheta^- G_{--}(\s) +\vartheta^+ p(\s) +\ri \vartheta^-\vartheta^+ \p_{--} g_+(\s)
~,
\eeqn
\esubeq
where $G_{\pm\pm}(\s)$ are the components of a vector current field
\be
\p_{++} G_{--}+\p_{--} G_{++}=0
~.
\ee

\subsection{$\mathcal N=(0,2)$ }\label{02flavor}
 
 Finally, the flavor current multiplet for $\cN=(0,2)$ supersymmetric theories is derived in 
  the appendix~ \ref{DeriveFlavorCurrent02}.   It contains two real superfields $\cG(\z)$ and $\cG_{--}(\z)$ satisfying the constraints:
   \be
\bar \cD_+(    \cG_{--}  -\ri  \p_\mm \cG )   =0
~, \qquad  
 \cD_+(    \cG_{--}  +\ri  \p_\mm \cG )  =0
 ~.
 \label{02flavorcurrent0}
\ee
These two equations are conjugate to each other. 

If we define the following descendant superfield 
\be
\cG_{++}= -\frac12 [\cD_+, \bar \cD_+] \cG
~,
\ee
then we have the conservation equation
\be\label{falvorCurrentNormal02}
\p_{++} \cG_{--}+\p_{--}\cG_{++}=0
~,
\ee
together with
\be
[\cD_+,\bar\cD_+]\cG_{--}= -2 \p_{++}\p_{--} \cG
~.
\label{D+D+G--02}
\ee 

The flavor current  multiplet is then described by the following decomposition in component fields
\bsubeq
\beqn
\cG(\z)&=&g(\s)+\ri \vartheta^+ p_+(\s) +\ri\bar\vartheta^+ \bar p_+(\s) +\vartheta^+\bar\vartheta^+ G_{++}(\s)
~,
\\
\cG_{--}(\z)&=&G_{--}(\s)+\vartheta^+ \p_{--} p_+(\s) -\bar\vartheta^+ \p_{--}\bar p_+(\s) 
+\vartheta^+\bar\vartheta^+ \p_{--}\p_{++} g(\s)
~,
\eeqn
\esubeq
where
\be
\p_{++} G_{--}+\p_{--} G_{++}=0
~,
\ee
 which is just the lowest component projection of    \eqref{falvorCurrentNormal02},
and indicates, once more, that $G_{\pm\pm}(\s)$ are the components of a vector current field.


\section{Supersymmetric $J\bar{T}$ and $T\bar{J}$   primary operators} \label{sec:TJbarsusy}

Let us first recall that, in light-cone notation, the standard   $T\bar{J}$ and $J\bar{T}$ composite  operators are defined as
\cite{Guica:2017lia}
\bsubeq
\bea \label{nonsusyTJbar}
O_{--}^{T\bar{J}}(\s)&:=&
T_{----}(\sigma)G_{++}(\s)
-\Theta(\sigma)G_{--}(\s)
~,
\\   \label{nonsusyJTbar}
O_{++}^{ J  \bar{T}}(\s)
&:=&
T_{++++}(\sigma)G_{--}(\s)
-\Theta(\sigma)G_{++}(\s)
~.
\eea
\esubeq
These two operators may be quite different in theories that are not parity invariant. This will indeed be the case
for theories with chiral supersymmetry, such as $\cN=(0,1)$ and $\cN=(0,2)$, that we are going to consider in our paper.

As already emphasized, $T\bar{J}$ and $J\bar{T}$ deformations break Lorentz invariance. 
This implies that the 
 stress-energy tensor is not symmetric anymore, $T_{++--}\neq T_{--++}$. 
Hence, the component $\Theta$ in the above two equations has two different meanings:   in \eqref{nonsusyTJbar}, 
$\Theta=T_{++--}$, while in  \eqref{nonsusyJTbar}, $\Theta=T_{--++}$
where the latter was defined as $\tilde{\Theta}$ in \eqref{Theta-ThetaTilde}. 
As already mentioned before, 
since $T_{++--}$,  $T_{--++}$ never appear simultaneously, in the following analysis we can forget about 
tildes.  We only need to make sure that the correct $\Theta$ is used and satisfies the
appropriate  conservation equations. 
 
 In this section, we will show that the $O_{--}^{T\bar{J}}$ and $O_{++}^{J\bar{T}}$ operators preserve supersymmetry
 in complete analogy with the $T\bar{T}$ case of \cite{Baggio:2018rpv,Chang:2018dge,Jiang:2019hux}. 
 More precisely, we will make use of the stress-tensor multiplets and flavor current multiplets introduced in the previous section
  to  construct supersymmetric primary $J\bar T$ and  $T\bar J$ operators
  and show that
the  $O_{--}^{T\bar{J}}$ and $O_{++}^{J\bar{T}}$ operators are supersymmetric descendants
of the primary ones
(up to total derivatives and equations of motion). 
Note that in the appendix~\ref{wellDef} we  discuss   well-definedness properties 
of all the supersymmetric primary operators
given in our paper. 
Interestingly,  
as already indicated in the introduction and elaborated in more detail in appendix \ref{SZgeneralization},
 it turns out that all the primary operators 
 fit into a general pattern
  which extends the original analysis of
\cite{Smirnov:2016lqw} and the supersymmetric extensions of \cite{Baggio:2018rpv,Chang:2018dge,Jiang:2019hux}.

\subsection{$\cN=(0,1)$}

  In section~\ref{01stressTensor} and section~\ref{01flavor}, we have presented the structure of the 
  $\cN=(0,1)$ stress-tensor multiplet 
  and the $\cN=(0,1)$ flavor current multiplet as well as their conservation equations.  
  With these ingredients, we can immediately construct two different bilinear  superfields 
  that work as supersymmetric primary operators for $T\bar{J}$ and $J\bar{T}$ in \eqref{nonsusyTJbar}.
They  are
\bsubeq
\bea \label{01JTbar}
\cO_+^{J\bar{T} }(\z)
&:=&
\cJ_{+++}(\z)\cG_{--}(\z)
-\cJ_-(\z)\cG_{++}(\z)
~,
\\  \label{01TJbar}
\cO_{---}^{{T}\bar{J}}(\z)
&:=&
\cT_{----}(\z)\cG_{+}(\z)
-\cJ_-(\z)\cG_{--}(\z)
~.
\eea
\esubeq
From these, it is in fact possible to construct the manifestly supersymmetric operators described by the following descendants
\bsubeq
\be
{\sf  O}_{++}^{J\bar{T} }(\s)
=\int\rd\vartheta^+\,\cO_{+}^{J\bar{T} }(\z)
~,
\ee
and
\begin{equation}
{\sf  O}_{--}^{T\bar{J}}(\s)
=\int\rd\vartheta^+\,\cO_{---}^{{T}\bar{J}}(\z)
~.
\end{equation}
\esubeq
These, up to conservation equations and total derivatives, prove to be equivalent to the $O_{--}^{T\bar{J}}$ and $O_{++}^{J\bar{T}}$
operators. 
As explained in more details for the $T\bar{T}$ case in \cite{Baggio:2018rpv},
this equivalence defines precisely how ${T\bar{J}}$ and ${J\bar{T}}$ deformations preserve 
$\cN=(0,1)$ supersymmetry. Similar results will hold for the $\cN=(1,1)$ and $\cN=(0,2)$ cases.

Let us start with the $J\bar T$ primary operator~\eqref{01JTbar}.  
One can straightforwardly  compute its descendant and obtain the following result
\bea
\cD_+\cO_+^{J\bar{T}  }
&=&
\cT_{++++}\cG_{--}
-\cT\cG_{++} 
+\ri \big(\p_{--}\cJ_{+++}+\p_{++}\cJ_-\big)\cG_{+}
+\cJ_- \big( \cD_+\cG_{++}+\ri\p_{++}\cG_{+} \big)
\non\\&&
-\cJ_{+++} \big( \cD_+\cG_{--}- \ri\p_{--}\cG_{+} \big)
-\ri\p_{--}\big(\cJ_{+++}\cG_{+}\big)
-\ri\p_{++}\big(\cJ_-\cG_{+}\big)
~.
\label{01JTbar-descendant}
\eea
It is easy to recognize that the quantities in the first three brackets are exactly the conservation equations of 
the  stress-tensor and 
flavor current multiplets while the last two terms are just total derivatives
that do not contribute once one integrates over the $\s^{\pm\pm}$ bosonic coordinates.
On the other hand, the lowest $\vartheta^+=0$ component of the first two terms in \eqref{01JTbar-descendant}
are precisely the $J\bar{T}$ operator, $O_{++}^{J\bar{T}}(\s)$.
Therefore, up to total derivatives and equations of motion,
the descendant of the primary operator~\eqref{01JTbar} is exactly the standard $T\bar J$ operator in~\eqref{nonsusyTJbar}:
\be
{\sf O}_{++}^{J\bar{T} }(\s)
=
\cD_+\cO_+^{J\bar{T}  }(\z)|_{\vartheta^+=0}
+ \textrm{total derivatives}
=
O_{++}^{J\bar{T} }(\s)
+ \textrm{EoMs} + \textrm{total derivatives}
~.
\ee
Here ``EoMs'' means  those quantities which vanish once the equations of motion, or more 
precisely the conservation equations, are used.

For the $T\bar J$ case one can similarly compute
\beqn
\cD_+\cO_{---}^{{T}\bar{J}}&=&
\cT_{----}\cG_{++}-\cT\cG_{--}
+
\big(\cD_+\cT_{----} - \ri\p_{--}\cJ_{-} \big)\cG_{+}
+\cT_{----}\big(\cD_+\cG_+ -\cG_{++} \big)  
\non
\\
& &
+\cJ_-\big( \cD_+\cG_{--} -\ri \p_{--}\cG_+ \big) 
+\ri\p_{--}\big(\cJ_{-}\cG_{+}\big)
~.
\qquad
\eeqn
Again, the quantities in the first three brackets are exactly the conservation equations, thus vanish on-shell.
Therefore, exactly as in the $J\bar{T}$ case, it holds
\begin{equation}
{\sf O}_{--}^{T\bar{J}}(\s)
=
O_{--}^{T\bar{J}}(\s)
+\text{EoMs}+ \textrm{total derivatives}
~.
\end{equation}

Remember that Smirnov and Zamolodchikov, 
by extending the analysis by Zamolodchikov for $T\bar T$  deformations \cite{Zamolodchikov:2004ce},
have proven that, given any pairs of currents $(A_s,\,B_{s+2})$ and $(A'_{s'},\,B'_{s'-2})$
satisfying the conservation equations
\bea
\partial_{++} A_{s} &= - \partial_{--} B_{s+2}~,\qquad
\partial_{--} A'_{s'} &= - \partial_{++} B'_{s'-2}~,
\eea
where $s$ and $s'$ label the spins of the operators, 
then the following bilinear operators
\bea
O^{\rm SZ}_{s+s'}(\s):=A_{s}(\sigma)\, A'_{s'}(\sigma)
-B_{s+2}(\sigma)\, B_{s'-2}(\sigma)~,
\label{AABB}
\eea
can be proven to be free of short distance
singularities and well defined by a point splitting procedure \cite{Smirnov:2016lqw}.
Both $O_{++}^{J\bar{T}}(\s)$ and $O_{--}^{T\bar{J}}(\s)$ are Smirnov-Zamolodchikov operators.
Note that the structure of $\cO_{+}^{\bar{T}J}(\zeta)$ \eqref{01JTbar} is the one of a Smirnov-Zamolodchikov type 
of operator given in \eqref{AABB}.
This implies that, exactly as the $\cN=(0,1)$ $T\bar{T}$ primary operator 
introduced in \cite{Baggio:2018rpv,Chang:2018dge},
 $\cO_{+}^{\bar{T}J}(\zeta)$, despite being a composite irrelevant operator, is free of short distance
singularities and well defined by a point splitting procedure as for the analysis in \cite{Smirnov:2016lqw}.
Interestingly, the $\cO_{---}^{{T}\bar{J}}(\z)$ is not of Smirnov-Zamolodchikov type.
Despite that,
as described in appendix~\ref{wellDef},
one can show that $\cO_{---}^{{T}\bar{J}}(\z)$
is also well defined,
in complete analogy to the analysis of \cite{Jiang:2019hux} where the $\cN=(0,2)$ $T\bar{T}$ operator was shown to be well defined 
even though not being of Smirnov-Zamolodchikov type.%
\footnote{See also \cite{Chang:2019kiu} for the $\cN=(2,2)$ case
which is also described by a $T\bar{T}$ primary operator that is not of Smirnov-Zamolodchikov type.}

\subsection{$\cN=(1,1)$}

From the stress-tensor multiplet and flavor current multiplet in subsections~\ref{11stressTensor} and \ref{11flavor}, 
we can construct the following primary operator 
\be
\cO_{++}^{J\bar{T}}(\z)=\cJ_\ppp(\z) \cG_-(\z)  + \cJ_+(\z) \cG_+(\z)
~.
 \ee
By using conservation equations, a straightforward calculation gives 
\be
\cD_-\cD_+ \cO_\pp  = \cT_{++++}  \cG_\mm  - \cT \cG_\pp +\text{total derivatives }+\text{EoMs}
~.
\ee
In complete analogy to the $\cN=(0,1)$ case of the previous subsection, this result implies
that the $J\bar{T}$ operator is equivalent to the descendant of the operator $\cO_\pp^{J\bar{T}}$, 
up to conservation equations and total derivatives:
\be
O_{++}^{J\bar{T} }(\s)
=\int\rd\vartheta^-\rd\vartheta^+\,\cO_{++}^{J\bar{T} }(\z)
+ \textrm{EoMs} + \textrm{total derivatives}
~.
\ee
 
For the $T\bar{J}$ case we can construct the following primary operator
\be
\cO_{--}^{T\bar{J} }(\z)=\cJ_\mmm(\z) \cG_+(\z)  + \cJ_-(\z) \cG_-(\z)
~.
 \ee
In complete analogy to all the other cases considered so far, one can prove the equivalence of its descendant operator
with the $ O_{--}^{T\bar{J} }(\s)$:
\be
 O_{--}^{T\bar{J} }(\s)
=\int\rd\vartheta^-\rd\vartheta^+\,\cO_{--}^{T\bar{J} }(\z)
+ \textrm{EoMs} + \textrm{total derivatives}
~.
\ee
 
Note that, since $\cN=(1,1)$ supersymmetry is left-right symmetric, the two $T\bar J$ and $J\bar T$ 
primary operators above are simply related through a parity transformation 
which exchanges the left and right moving sectors. 

To conclude this subsection, note also that in the $(1,1)$ case both these operators are not 
of Smirnov-Zamolodchikov type.
Despite that,
as described in appendix~\ref{wellDef}, it is once more possible to use the arguments originally presented in 
\cite{Jiang:2019hux} and show that 
$T\bar J$ and $J\bar T$ 
primary operators are both well defined.

\subsection{$ \cN=(0,2)$}

Finally we can turn to discuss  the $T\bar J/J\bar T$ supersymmetric primary operators in $\cN=(0,2)$ theories
that are constructed as bilinears of the stress-tensor multiplet and flavor current multiplet 
 given in subsection~\ref{02stressTensor} and \ref{02flavor}.
The well-definedness of  various operators is analyzed in appendix~\ref{wellDef}.

  \subsubsection*{$\bullet$ \large $\;  \cN=(0,2)\; J\bar T$}

For the $J\bar T$ case, we can naturally construct the following primary operator
\be\label{02JTbar-0}
\cO ^{J\bar{T} }(\z)=\cS_{++}(\z)\cG_{--}(\z) - 2\cG(\z) \cT(\z)
~.
\ee
It is easy to check that it holds
\be
 \frac14[ \bar \cD_+, \cD_+]\cO ^{J\bar{T} }(\z)= \cT_{++++}(\z)  \cG_{--}(\z)-\cT(\z)  \cG_{++}(\z) +\text{EoMs}+\text{total derivatives}
 ~,
\ee
which implies
\be
 O_{++}^{J\bar{T} }(\s) =\frac12 \int\rd\bar\vartheta^+\rd\vartheta^+\,\cO  ^{J\bar{T} }(\z)
+ \textrm{EoMs} + \textrm{total derivatives}
~,
\label{descendant-02JTbar}
\ee
as expected.

By remembering from eq.~\eqref{descendants-S-multiplet} that
$\cT:=\frac{\ri}{2} \big( \cD_+W_- -\bar \cD_+\bar W_- \big)$,
it is clear that, up to terms that are $\cD_+$ and/or  $\bar\cD_+$ acting on a superfield, 
the second term in \eqref{02JTbar-0} can be written in different equivalent ways while preserving the main result
\eqref{descendant-02JTbar}.
In fact,  for $\cN=(0,2)$ theories with an $R$-symmetry there is a very natural variant definition of the supersymmetric primary
operator in terms of the $\cR$-multiplet superfields $\cR_{\pm\pm}$. 
This is given by the following operator
\be
\cO_{\cR}^{J\bar{T} }(\z)=\cR_{++}(\z)\cG_{--}(\z) -  \cR_{--}(\z) \cG_{++}(\z)
~,
\label{02JTbar-R}
\ee
which is such that
$\cO_{\cR}^{J\bar{T} }(\z)=\cO^{J\bar{T} }(\z)+ \cD_+ (\cdots)+ \bar\cD_+ (\cdots)$
and clearly also satisfies eq.~\eqref{descendant-02JTbar}.
Note that $\cO_{\cR}^{J\bar{T} }(\z)$ in \eqref{02JTbar-R} is of Smirnov-Zamolodchikov type.

 \subsubsection*{ $\bullet$ \large$\;\cN=(0,2)\; T\bar J$}

For simplicity, in the $T\bar J$ case let us start directly from an $\cN=(0,2)$ supersymmetric theory possessing an $R$-symmetry.
After that, we will extend the $\cR$-multiplet results to the general case in which the stress-tensor multiplet is 
an $\cS$-multiplet.

Assuming the existence of an $\cR$-multiplet, we can construct the following supersymmetric primary operator
\be\label{TJbar02R}
\cO_{----}^{T\bar J}(\z)=\cT_{----}(\z)\cG(\z)-\frac12 \cR_{--}(\z) \cG_{--}(\z)
~.
\ee
In analogy to the other supersymmetric primary operators considered so far, a straightforward calculation
leads to the following result
\be
\frac12 [\bar \cD_+, \cD_+]\cO_{----} =\cT_{----}  \cG_{++}-\cT \cG_{--} +\text{EoM}+\text{total derivatives}
~,
\ee
which implies
\be
 O_{--}^{T\bar{J} }(\s) =  \int\rd\bar\vartheta^+\rd\vartheta^+\,\cO_{----}^{T\bar{J} }(\z)
+ \textrm{EoMs} + \textrm{total derivatives}
~,
\label{descendantTJbar02}
\ee
as expected.

In  the absence of a conserved $R$-symmetry in the stress-tensor multiplet, 
one can not construct the  $T\bar J$ supersymmetric primary operator 
as a unique $D$-term whose full superspace integral leads to $O_{--}^{T\bar J}$.
The reason is simply that there might not exist in the $\cS$-multiplet the $\cR_{--}$ operator such that 
$\cW_-=\frac{\ri}{2}\bar{\cD}_+\cR_{--}$
and $\cW_-=\frac{\ri}{2}\bar{\cD}_+\cR_{--}$, eq.~\eqref{calW=DR}.
In such a case, $\cO_{----}^{T\bar J}$ of eq.~\eqref{TJbar02R} will not exist and consequently
eq.~\eqref{descendantTJbar02} will not hold.
Nevertheless, for a general $\cN=(0,2)$ supersymmetric theory 
it is still possible to show that $O_{--}^{T\bar{J} }(\s)$ arises as a linear combination
of a full superspace integral and a chiral half superspace integral.%
\footnote{Note that the same happens with $T\bar{T}$ deformations for general $\cN=(2,2)$ supersymmetric 
models described by an $\cS$-multiplet \cite{Chang:2019kiu} where the $T\bar{T}$ operator 
is related to a linear combination of full and chiral superspace integrals.} 
This extends eq.~\eqref{descendantTJbar02}
and proves again that ${T\bar{J} }$ deformations preserve supersymmetry.
Let us turn to the precise description of this case.

First, note that the constraints defining the $\cN=(0,2)$ flavor current multiplet, eq.~\eqref{02flavorcurrent0},
tell us that the superfields
\be
\cH_{--}   :=  \cG_{--}  -\ri  \p_\mm \cG 
~,\qquad 
\bar\cH_{--}   =  \cG_{--}  +\ri  \p_\mm \cG  
~,
\ee
are chiral and anti-chiral, respectively:  $\bar \cD_+ \cH_{--}=0$,  $  \cD_+\bar \cH_{--}=0$.
Moreover, they satisfy the following relations
\be
\cD_+ \cH_{--} =-2\ri \p_{--}\cD_+ \cG
~, \qquad   
\bar \cD_+ \bar \cH_{--} = 2\ri \p_{--} \bar \cD_+ \cG
~.
\ee
By using  $\cH_{--}$ and $\bar{\cH}_{--}$ together with the superfields of the $\cS$-multiplet, 
we can define the following $T\bar J$ superfield
\be
\mathscr  O_{--}^{T\bar J}=\frac12 [\bar \cD_+ , \cD_+  ]  \big( \cT_{----}\cG   \big)
-  \frac{\ri}{2}     \cD_+  \big( \cW_-   \cH_{--}  \big) 
+\frac{\ri}{2} \bar\cD_+ \big(\bar \cW_- \bar\cH_{--}\big)
~.
\ee
This can be easily shown to be
\be
\mathscr   O_{--}^{T\bar J}=\cT_{----}  \cG_{++}-\cT \cG_{--} +\text{EoM}+\text{total derivatives}
~,
\ee
whose   lowest $\vartheta=0$ component is just the standard $T\bar J$-operator
\bea
 O_{--}^{T\bar J}
 &=&
 \int \rd \bar \vartheta^ +  \rd\vartheta^+ \cT_{----}\cG  
 -  \frac{\ri}{2}   \Big(    \int \rd   \vartheta^ +  \cW_-   \cH_{--}  -\int \rd \bar \vartheta^ + \bar \cW_- \bar\cH_{--}\Big)
 \non
 \\
&& +\text{EoM}+\text{total derivatives}
~,
\eea
as expected.

 
\section{Examples of supersymmetric $J\bar T/T\bar J$ deformations} \label{TJexample}

In this section, we will present some explicit examples of  supersymmetric $J\bar T/T\bar J$ deformations. 
 As argued in \cite{Guica:2017lia}, $J\bar T/T\bar J$ is solvable   when the U(1) current is chirally conserved. 
With the aim of extending the results of \cite{Guica:2017lia}, in this section
we will only focus on supersymmetric examples arising from \emph{chiral} $J\bar T/T\bar J$ deformations.
Our analysis will be purely classical here but we will manage to construct explicit $J\bar T/T\bar J$ flows for some simple 
supersymmetric example.

In both the $\cN=(0,1)$ and $\cN=(0,2)$ cases, 
we will present two models induced by     $J\bar T$ and $T\bar J$ deformations, respectively.
The chiral   $J\bar T/T\bar J$ deformations  in $\cN=(1,1)$ theories
seem  to resist illustrations in simple examples. We will comment more on these cases
in the conclusion.

\subsection{$\cN=(0,1)$   $J\bar T$} \label{01JTbarExample}

Here we are going to present the simplest example of $J\bar T$ deformation with  $\cN=(0,1)$   supersymmetry. 
 It consists of a  left-moving complex 
fermion which has the U(1) symmetry, and  a right-moving  supersymmetric  sector which consists of a real scalar and a real fermion.

  \subsubsection{Component form}

Inspired by the non-supersymmetric example in \cite{Guica:2017lia}, we propose that  the following action 
satisfies a $J\bar T$ flow. 
\bsubeq
\beqn\label{bosdef01}
S_{\alpha}&=&\int \rd^2 \sigma \mathcal L_\alpha 
=\int \rd^2 \sigma  \Big(\mathcal L_L+\mathcal L_R+\alpha\mathcal L_\text{def} \Big) 
~,\\
\mathcal L_L &=& \frac{\ri}{2} \bar \chi_- \p_{++} \chi_-    \label{01JTbarLeft}
~, \\ 
\mathcal L_R &=& \frac12 \p_{++} \phi \p_{--}\phi +\frac{\ri}{2} \psi_+ \p_{--} \psi_+    \label{01JTbarRight}
~,\\ 
\mathcal L_\text{def} &=&  - \bar \chi_- \chi_- \Big(   \p_{++} \phi \p_{++}\phi + \ri \psi_+ \p_{++} \psi_+  \Big)  
~.
\eeqn
\esubeq
where $\chi_-$ is a complex fermion, while $\psi_+=\bar\psi_+$ and $\phi=\bar \phi$ 
are a real fermion and a real scalar, respectively. 

The left-moving complex fermion possess the following U(1) symmetry  
\be
\chi_-\rightarrow e^{\ri\rho}\chi_-
~, \qquad   \bar \chi_-\rightarrow e^{-\ri\rho} \bar \chi_-
~.
\label{chiral01sym}
\ee
According to Noether's theorem, this gives rise to the following  U(1) current 
\be
G_{--}=\bar \chi_- \chi_-
~,
\ee
which we will shortly show to be chiral.

The $T_{++++}$ component of the stress-energy tensor for the action \eqref{bosdef01}
 can be easily computed and turns out to be~%
\footnote{Our conventions for the  stress-energy tensor is 
$T_{ab} =\eta_{ac} \frac{\p \mathcal L}{\p \p_c \varphi} \p_b\varphi -\eta_{ab} \mathcal L$ 
where in light-cone notations 
the Minkowski metric is
$\eta_{\pm\pm, \pm\pm}=-2,\eta^{\pm\pm, \pm\pm}=-\frac12, \eta_{\pm\pm, \mp\mp}=\eta^{\pm\pm, \mp\mp}=0$.
 }
\be
T_{++++}=- \p_{++} \phi \p_{++}\phi -\ri \psi_+ \p_{++} \psi_+
~.
\ee
In particular, note that  $\mathcal L_\text{def}$ does not contribute to this component. 

Note that the equation of motion of the complex fermion  $\chi_-$ is
\be\label{EoMchi}
\p_{++}\chi_-=-2\alpha\ri \chi_-   \Big(   \p_{++} \phi \p_{++}\phi + \ri \psi_+ \p_{++} \psi_+  \Big) =2\alpha\ri \chi_- T_{++++}
~.
\ee
Together with its complex conjugate, 
a short but instructive calculation which uses \eqref{EoMchi}
shows that  it holds
\be
\p_{++} G_{--}=0
~.
\ee
Therefore, the U(1) current is chirally conserved and $G_{++}=0$,  
which is expected from the symmetry \eqref{chiral01sym} 
where there is even no notion of  $G_{++}$.

Finally, we easily notice that the deformation part of the Lagrangian \eqref{bosdef01} satisfies (remember that $G_{++}=0$)
\be
\frac{\p  \mathcal L_\alpha}{\p  \alpha} =\mathcal L_\text{def}  =T_{++++} G_{--}
=O^{J\bar T}
~.
\ee
This shows that the action we proposed in eq.~\eqref{bosdef01}
arises from a $J\bar T$ deformation as expected.
Since the U(1) current is chirally conserved, the deformation  is thus a \emph{chiral} $J\bar T$ deformation. 

So far we have not discussed whether the model described by \eqref{bosdef01} is supersymmetric, 
though from our general discussion
we expect this to be the case.
To prove this statement explicitly we turn to describing the same model in $\mathcal N=(0,1)$ superspace.

 \subsubsection{Superfield form}

 We start by introducing the following superfields 
\be
\Phi =\phi -\ri \vartheta^+ \psi_+
~,
\label{Phi01}
\ee
and
\be
\Upsilon_-=\chi_--\ri\vartheta^+ B
~,~~~~~~
\bar{\Upsilon}_-=\bar\chi_-+\ri\vartheta^+ \bar{B}
~,
\label{Upsilon01}
\ee
 that embed the component fields 
 $\phi$, $\psi_+$, $\chi_-$ and $\bar\chi_-$ into appropriate supermultiplets.

A natural manifestly supersymmetric extension of the action \eqref{bosdef01} is
 \bea
S_\alpha &=&  \int \rd^2 \sigma    \rd\vartheta^+  \Big(
\frac{\ri}{2} \cD_+ \Phi \p_{--}\Phi 
   +  \frac{1}{2} \bar{\Upsilon}_-\cD_+ \Upsilon_-
- \ri\alpha \bar \Upsilon_- \Upsilon_- \cD_+ \Phi \p_{++}\Phi  \Big)
~.
\label{JTdef01super}
\eea
To show the equivalence with  \eqref{bosdef01}, 
 we reduce \eqref{JTdef01super} to components and obtain
\bea
S_\alpha 
&=& 
\int \rd^2 \sigma \;  \Bigg[  ~
  \frac12 \p_{++} \phi \p_{--}\phi +\frac{\ri}{2} \psi_+ \p_{--} \psi_+    
 +\frac{\ri}{2}   \bar \chi_- \p_{++} \chi_-     
+\frac12 B\bar B
\nonumber\\&&\qquad\qquad
-\ri \alpha ( \chi_- \bar B+\bar \chi_- B )\psi_+\p_\pp \phi
 - \alpha\bar \chi_- \chi_- \Big(   \p_{++} \phi \p_{++}\phi + \ri \psi_+ \p_{++} \psi_+  \Big)   
 \Bigg]
 ~.~~~~~~
 \label{JTdef01super-2}
\eea
Note that the previous action is identical to \eqref{bosdef01} except for all the terms involving 
the complex auxiliary fields $B$ and $\bar B$.
It is simple to show that these terms are identically zero once we integrate out $B$ and $\bar B$.
In fact, these can be solved in terms of the physical fields by using their algebraic equation of motion:
\beqn\label{aux01}
B=2\ri\alpha \chi_- \psi_+ \p_{++}\phi
~, \qquad 
\bar B=2\ri\alpha \bar \chi_- \psi_+ \p_{++}\phi
~.
\eeqn
By substituting this result  back into \eqref{JTdef01super-2}, one can see that the auxiliary fields 
$B$ and $\bar B$ have no contribution  due to  the fermionic property $\psi^2_+=0$. 
Thus the manifestly supersymmetric action \eqref{JTdef01super}--\eqref{JTdef01super-2}
is equivalent to the $J\bar T$ deformed action \eqref{bosdef01}.
The above construction also tells us  that   the  action \eqref{bosdef01} is supersymmetric. 

To see the supersymmetry more explicitly, we can work out the supersymmetry transformation rules.   
The off-shell $\cN=(0,1)$ supersymmetry transformation of an arbitrary superfield $\cF$
was given in \eqref{01susyTsf} and we repeat them here for the reader's convenience:
\be
\delta \mathcal F=-\ri \epsilon_- \mathcal Q_+ \mathcal F 
=-\ri \epsilon_- \Big( \ri\frac{\p}{\p \vartheta^+ }  -\vartheta^+ \p_{++} \Big)  \mathcal F
~.
\ee
By using this rule for $\Phi$,  eq.~\eqref{Phi01}, and $\Upsilon_-$, eq.~\eqref{Upsilon01},
one can derive the off-shell supersymmetry transformations of their component fields
\be
\delta \chi_- =-\ri \epsilon_- B
~, \qquad 
\delta \bar \chi_- = \ri \epsilon_- \bar  B
~, \qquad 
\delta \phi =-\ri \epsilon_- \psi_+
~, \qquad 
\delta \psi_+=\epsilon_-\p_{++}\phi
~.
\label{01susy-components}
\ee
One can check explicitly that \eqref{JTdef01super-2} is invariant under the previous transformations.
Note also that the equations of motion for $B$ and $\bar B$ given by \eqref{aux01} are also consistent with these supersymmetry 
transformations. 
In fact, one can also verify that
\eqref{bosdef01} is 
 invariant under \eqref{01susy-components} \emph{on-shell}, meaning when \eqref{aux01} are satisfied. 
In particular, one can check that in this case $\delta (\bar \chi_- \chi_-) =0$
 which guarantees that no higher-order terms in $\alpha $ would be generated in the on-shell transformation rules.   
We can then conclude that the model described by the action \eqref{bosdef01}, which we have previously shown to be 
a standard $J\bar T$ deformation, is  supersymmetric as expected.

Let us look back at the manifestly off-shell supersymmetric action \eqref{JTdef01super}
and show that it is a manifestly supersymmetric deformation associated to the operator \eqref{01JTbar}.
First, we rewrite the action \eqref{JTdef01super} as
   \beqn\label{01superaction}
S_\alpha &=&  \int \rd^2 \sigma    \rd\vartheta^+ \mathcal A_\alpha
~, 
\quad \mathcal A_\alpha= \Big(
\frac{\ri}{2} \cD_+ \Phi \p_{--}\Phi 
   +  \frac{1}{2} \bar{\Upsilon}_-\cD_+ \Upsilon_-
- \ri\alpha \bar \Upsilon_- \Upsilon_- \cD_+ \Phi \p_{++}\Phi  \Big) 
~.~~~~~~~~~
\eeqn
We can derive the stress-tensor multiplet for example by using the Noether procedure of \cite{Chang:2018dge}.
We obtain
\be
\cJ_{+++}= - \ri \frac{\delta \mathcal A_\alpha}{\delta \p_{--}\Phi}\p_{++}\Phi
=- \ri  \cD_+\Phi \p_{++}\Phi
~.
\ee
Then it is easy to see that the superspace Lagrangian $\cA_\a$ satisfies the supersymmetric $J\bar T$ flow equation
\be\label{01superflow}
\frac{\p \mathcal A_\alpha }{\p \alpha} =\cJ_{+++}\cG_{--}
=\cO^{J\bar{T}}_+
~,
\ee
where
\be\label{01Gmm}
\cG_{--}=\bar \Upsilon_- \Upsilon_-= \bar\chi_- \chi_- +\ri \vartheta^+ (\bar \chi_- B+\chi_- \bar B)
~,
\ee
and $\cG_{++}=0$ in 
the supersymmetric primary operator $\cO^{{J\bar{T}}}_+$ of eq.~\eqref{01JTbar}.
Let us in fact verify at the superspace level that $\cG_{--}$ is chirally conserved.
By using the superspace equations of motion for the superfields $\Upsilon_-$ and $\bar \Upsilon_-$
which read
\be
\cD_+ \Upsilon_-= 2\ri \alpha \Upsilon_- \cD_+\Phi \p_{++}\Phi
~, \qquad
\cD_+\bar  \Upsilon_-=- 2\ri \alpha \bar\Upsilon_- \cD_+\Phi \p_{++}\Phi
~,
\ee
it is a straightforward calculation to prove the following result
 \be\label{01DGmm}
 \cD_+ \cG_{--}
 =
 \cD_+\bar  \Upsilon_- \cdot \Upsilon_- 
 - \cD_+  \Upsilon_-  \cdot \bar  \Upsilon_-
 =- 2\ri \alpha (\bar  \Upsilon \Upsilon_-  +   \Upsilon\bar \Upsilon_- ) \cD_+\Phi \p_{++}\Phi=0
 ~.
 \ee

Note that the conservation equation~\eqref{01DGmm} 
 is expected considering that the action~\eqref{01superaction} is invariant under the following  symmetry 
\be
\Upsilon_-\rightarrow e^{\ri\rho}\Upsilon_-
~, \qquad  
 \bar \Upsilon_-\rightarrow e^{-\ri\rho} \bar \Upsilon_-
 ~.
 \label{01sym2}
\ee
Actually, the super flavor current~\eqref{01Gmm}  can also be constructed directly by promoting~\eqref{01sym2}
to a gauge symmetry and then covariantizing~%
\footnote{More specifically, by using the covariant derivative $\nabla_A$ in \eqref{covD01} in place of $\cD_A$.} 
the action~\eqref{01superaction}. 
Comparing the linearized action with \eqref{01gaugecopuling} gives \eqref{01Gmm} and $\cG_+=0$.

To summarize, we have shown that the superspace action~\eqref{01superaction} 
arises from a $J\bar T$ deformation with $\cN=(0,1)$ supersymmetry, and satisfies the 
manifestly supersymmetric $J\bar T$ 
flow equation \eqref{01superflow} driven by the supersymmetric primary operator $\cO^{J\bar{T}}_+$ 
of eq.~\eqref{01JTbar}.

 \subsection{$\cN=(0,1)$   $T\bar J$}
\label{TJbar01}

In this subsection, we are going to present an $\cN=(0,1)$ supersymmetric model which arises from the chiral $T\bar J$ deformation. 
This is going to be a supersymmetric generalization of the model first presented by Guica in \cite{Guica:2017lia}.
We start by reconstructing this model in the case without supersymmetry and then turn to its $\cN=(0,1)$ supersymmetric
extension.

\subsubsection{A   bosonic $T\bar J$ model from a new perspective  }

In \cite{Guica:2017lia}, Guica worked out the $T\bar J$-deformation of a free scalar field action. The U(1) current is associated with 
the shift symmetry of the  real 
free massless scalar field. Here we would like to rederive this model from a slightly different point of view
which will be used in constructing the supersymmetric extension. 

We can make the following educated guess for the action of the $T\bar J$-deformed real free massless scalar field
\be\label{TJdefactionboson}
S_\l=\int \rd^2\sigma  \;  \mathcal L_\lambda =\int \rd^2\sigma  \; \p_{++} \phi \p_{--} \phi   \,
F(\lambda \p_{--}  \phi )
~.
\ee
Here $F(x)$ is an arbitrary analytic function such that $F(0)=1$, which ensures that the undeformed action, $S_0$, is the one
of a free massless scalar field. 
 The $ T_{----}$ component of the stress-energy tensor
of the action \eqref{TJdefactionboson} proves to be 
\be\label{Tmmmm}
 T_{----}=-2(\p_{--} \phi )^2  F
 ~.
\ee

We want to impose the action \eqref{TJdefactionboson} to have a conserved chiral current
$G_{++}$, which is $\p_{--}G_{++}=0$, and to be a  $T\bar J$ flow, namely:
\be\label{TJdefbos}
\p_\lambda \mathcal L_\lambda= G_{++}  T_{----} 
~.
\ee
As already mentioned, the reason to consider 
a chiral $T\bar J$-deformation is that this is the case for which the quantum 
spectrum of the model is still solvable \cite{Guica:2017lia}.  
The  $T\bar J$ flow equation \eqref{TJdefbos} together with \eqref{Tmmmm} can be used to determine 
the  U(1) current
\be\label{Gppboson}
G_{++}=-   \frac{F'}{2F}\p_{++} \phi  
~.
\ee
The previous result is consistent only when we assume $G_{++}$ to be chirally conserved on-shell, 
which turns into the following 
constraint
\be
 \p_{--} G_{++}=\p_{--} \Big(-   \frac{F'}{2F}\p_{++} \phi  \Big)=0
 ~.
\ee
Using the equation of motion for the action \eqref{TJdefactionboson}, 
the above conservation equation leads to the following differential equation for the function $F(x)$:
\be\label{diffFbos}
\frac{F'+\frac12 xF''}{F+x F'}=\frac{F''}{F'}-\frac{F'}{F}
~, \qquad x=\lambda \p_{--} \phi 
~.
\ee
Solving this equation,  one gets
\be
F=\frac{c_2}{x+c_1}
~, \qquad  \text{or} \quad 
F=\frac{c}{x^2}
~.
\ee
Once we impose the boundary condition $F(0)=1$,
the second solution is discarded and the most general solution  turns out to be:
\be\label{solF}
F(x)=\frac{c}{x+c}
~.
\ee
By plugging this result into \eqref{Gppboson},
the chiral current is then given by
\be
G_{++}=-   \frac{F'}{2F}\p_{++} \phi  = \frac12 \frac{1}{c+\lambda \p_{--} \phi } \p_{++}\phi 
~.
\ee
In the undeformed limit $\lambda=0$, it holds $G_{++}=\frac{1}{2c} \p_{++} \phi $. 
Therefore, the seemingly extra parameter $c$  just corresponds to the normalization of the current which we have not specified yet 
and  has no physical meaning. 
To be consistent with \cite{Guica:2017lia}, we  choose the normalization  $c=-4$, hence the function $F(x)$ is
\be\label{solFmonica}
F(x)=\frac{1}{1-\frac14x }
~.
\ee

To conclude, we have shown that  the action \eqref{TJdefactionboson} 
with $F$ given by \eqref{solFmonica} describes a chiral $T\bar J$ flow,
eq.~\eqref{TJdefbos}, 
with chiral current given by \eqref{Gppboson}.

\subsubsection{A   $T\bar J$ deformed model with $\cN=(0,1)$ supersymmetry}

Now we would like to extend the analysis of the previous subsection to the supersymmetric case 
and find the $T\bar J$-deformation of a free $\cN=(0,1)$ scalar multiplet action. 
The natural 
manifestly off-shell supersymmetric extension of \eqref{TJdefactionboson} is  given by the following ansatz
\be\label{01TJbarsuperaction}
S_\l=\ri\int \rd^2 \sigma \rd\vartheta^+\; \cD_+ \Phi \p_{--} \Phi F(\lambda \p_{--}\Phi)
~,
\ee
where the real scalar superfield $\Phi(\z)$ is the same as \eqref{Phi01} and the analytic function $F(x)$ is such that $F(0)=1$
but otherwise arbitrary.
Similarly to the $\cN=(0,1)$ $J\bar T$ deformation, we will first analyze the previous ansatz in components and then directly
in superspace.

\subsubsection*{$\bullet $ \emph{ Component approach}}

Once the superfield $\Phi(\z)$ is reduced to its real component fields
$\phi(\s)$ and $\psi_+(\s)$, see eq.~\eqref{Phi01}, 
and the Grassmann integral is performed, the action \eqref{01TJbarsuperaction} takes the form
\beqn
S_\l=\int \rd^2\sigma\; \mathcal L_\lambda&=&
\int \rd^2\sigma\; \Big\{ \p_\pp \phi \p_\mm \phi  F+  \ri \psi_+ \p_\mm \psi_+\big( F +\lambda   \p_{--} \phi F' \big)\Big\}
~.
\label{01TJbaraction}
\eeqn
The $T_{----}$ component of the  stress-energy tensor for the previous model proves to be
\be
T_{----}=-2  (\p_\mm \phi )^2F
~.
\ee
Interestingly, this is exactly the same as the bosonic case \eqref{Tmmmm}.

As in the pure bosonic case,  we want to interpret the action \eqref{01TJbaraction} as a chiral $T\bar J$-deformation
satisfying   
\be\label{01TJbarflow}
\p_\lambda \mathcal L_\lambda=   G_{++} T_{----}
~.
\ee
This flow equation enables us to determine the U(1) current to be
\be\label{Jppsusy}
G_{++}
= - \frac{1}{2F}\Big(F'\p_\pp \phi +\ri \lambda F'' \psi_+\p_\mm \psi _+ \Big) 
- \ri   \frac{F'}{F}  \frac{\psi_+\p_\mm \psi_+ }{\p_\mm \phi   }
~.
\ee
Consistency of \eqref{01TJbarflow} requires that $G_{++}$ is chiral, $\pa_{--}G_{++}=0$, which we are going to study next.

The equation of motion for the fermion of the action \eqref{01TJbaraction}  is given by:
\beqn
0
&=&2\ri \p_\mm \psi_+  \cdot (  F  +  \lambda F'   \p_\mm \phi   )+\ri \psi_+\p_\mm (  F  +  \lambda F'   \p_\mm \phi   )
~,
\eeqn
which yields
\be
\p_\mm \psi_+= -  \psi_+ \frac{\p_\mm( F  + x F' )}{ 2(F+x F')}
~.
\ee
Multiplying by $\psi_+$, we get the following non-trivial simplification
\be\label{psiPpsi}
\psi_+\p_\mm \psi_+=0 
~.
\ee
Interestingly, the previous result implies 
that $G_{++}$, eq.~\eqref{Jppsusy}, 
has no contribution from the fermion $\psi_+$ once its equation of motion is used. In this case,
\eqref{Jppsusy} simplifies to
\be\label{Gppsimple01}
G_{++}= - \frac{ F'}{2F} \p_\pp \phi 
~,
\ee
which is precisely the same as the purely bosonic case, eq.~\eqref{Gppboson}.
Note also that by using \eqref{psiPpsi} the fermion terms disappear from the action \eqref{01TJbaraction}.
This implies that the dynamics of the boson $\phi$ can be treated independently from the fermion $\psi_+$,
once \eqref{psiPpsi} holds.
Therefore, for the purpose of imposing that $G_{++}$ is chiral on-shell,
effectively one can use eq.~\eqref{Gppsimple01} and note that $\phi$ has the same equation of motion
as for the purely bosonic action \eqref{TJdefactionboson}.
This immediately implies that the condition for $G_{++}$ to be a chiral current is solved by the same function $F$ as in the 
non-supersymmetric case, namely  \eqref{solF}. This concludes the proof that
the supersymmetric action  \eqref{01TJbaraction}, and equivalently \eqref{01TJbarsuperaction}, 
satisfies the $T\bar{J}$ flow \eqref{01TJbarflow} with a chirally conserved current 
$G_{++}$ given by \eqref{Jppsusy}.

\subsubsection*{$\bullet$ \emph { Superfield approach}} 

In the discussions above, we have worked out the $T\bar J$ deformation in terms of component fields. 
It is natural to expect that \eqref{01TJbarsuperaction} satisfies a 
$T\bar J$ flow equation driven by the superfield operator \eqref{01TJbar}.
We show this to be true in the following.

The action \eqref{01TJbarsuperaction}  is given by the superspace integral of the superfield Lagrangian $\cA_\l$:
\be\label{TJ01super}
S_\l= \int \rd^2 \sigma \rd\vartheta^+\;  { \mathcal {A}_\lambda  }
~, \qquad 
\mathcal A_\lambda=\ri \cD_+ \Phi \p_{--} \Phi F(\lambda \p_{--}\Phi)
~.
\ee
By using for example the Noether techniques for $\cN=(0,1)$ superspace described in \cite{Chang:2018dge},
one can compute the $\cT_{----}$ superfield component of the stress-tensor multiplet:
\be
\cT_{----}=2\ri\Bigg[ \frac{\delta \mathcal A}{\delta \cD_+ \Phi}\p_{--}\Phi 
-\ri\cD_+ \Big( \frac{\delta \mathcal A}{\delta \p_{++} \Phi} \p_{--}\Phi \Big) \Bigg]
=-2 (\p_{--} \Phi)^2 F(\lambda \p_{--}\Phi)
~.
\label{calT----0}
\ee
Note that its  lowest $\vartheta=0$ component gives 
 the corresponding component of the stress-energy tensor, $\cT_{----} |_{\vartheta=0}=T_{----}$, as expected. 

If the action \eqref{TJ01super} arises from a chiral $T\bar J$ deformation, 
it should satisfy the following flow equation
\be
\p_\lambda \mathcal A_\lambda =\cO_{---}^{T\bar J}=\cT_{----} \cG_+  
~,
\ee
with $\cG_{--}=0$.
Thus, by imposing the previous flow equation for the Lagrangian $\cA_\l$ in \eqref{TJ01super} and the expression for 
$\cT_{----}$ given by \eqref{calT----0}, the superfield $\cG_+$ can be solved as
\be
\cG_+= -\ri \cD_+ \Phi   \frac{ F'}{2 F} 
~.
\label{calG}
\ee
For consistency, with $\cG_{--}=0$,
$\cG_+$ should describe a chiral current multiplet satisfying~\eqref{flavorcurrent01}
\be
\p_{--} \cG_+=0
~.
\ee
By imposing this constraint on \eqref{calG} one obtains
\be
    \p_{--}\cD_+ \Phi  FF'+\lambda  \cD_+ \Phi \p_{--}^2  \Phi (FF''-F'^2)=0
    ~,
    \label{constrGchiral01}
\ee
which should hold on-shell. 
The superspace equation of motion for the real scalar superfield $\Phi$ can be easily computed
by varying the action \eqref{TJ01super} and is given by
\be\label{EoMsuper01}
2 G' \p_{--}\cD_+ \Phi +\lambda  \cD_+ \Phi \p_{--}^2  \Phi G''=0
~, \qquad G(x) =x F(x)
~,\qquad x=\lambda \p_{--} \Phi
~.
\ee
By using this result in \eqref{constrGchiral01} we can obtain the following equation 
\be
-\frac{\p_{--} \cD_+ \Phi}{\lambda \cD_+ \Phi \p_{--}^2 \Phi} 
=\frac{G''}{2G'}
=\frac{FF''-F'^2}{FF'}
~.
\ee
Using $G(x) =x F(x)$, we get 
\be
\frac{(2F+xF')(-2F'^2+FF'')}{2F F'(F+xF')}=0
~.
\ee
One can easily check that this differential equation is equivalent to the one we obtained in the bosonic case,
eq.~\eqref{diffFbos}. 
Thus the solution, of the above differential equation is also given by the bosonic one \eqref{solFmonica}. 
This is consistent with our previous component approach.

To make more clear the connection with the components results given above, we can further calculate~%
\footnote{Here we used the relation $F'=\frac14 F^2$ which holds for \eqref{solFmonica}. }
\bsubeq
\bea
\cG_{++}=\cD_+\cG_+
&=&
 -\frac18 \p_{++} \Phi F-\frac{\ri}{32} \lambda F^2 \cD_+ \Phi \p_{--} \cD_+ \Phi
~,
\\
&=& -\frac18 \p_{++} \Phi F
~,
\eea
\esubeq
where in the last equality we used the relation $\cD_+ \Phi \p_{--} \cD_+ \Phi =0$ which can be obtained by multiplying the equation 
of motion \eqref{EoMsuper01} with $\cD_+\Phi$. 
These can be seen to be in agreement with the components results for $G_{++}$ given above.
In particular, the  $\vartheta=0$  component projection of $\cG_{++}$ on-shell is given by
\be
G_{++}=\cG_{++}|_{\vartheta=0}=-\frac18 \p_{++} \phi  F
~,   
\ee
which is in agreement with \eqref{Gppsimple01}. In particular,  it follows that $\p_{--}G_{++}=0$ on-shell.

\subsection{$\cN=   (0,2)$   $J\bar T$}

In this subsection, we are going to generalize the model constructed for the $\cN=(0,1)$ case
by complexifying  its right-moving sector. 
 The resulting model 
will possess the left-moving complex fermion $\chi_-$ and $\bar{\chi}_-$, which generates the U(1) symmetry, 
and  a 
complex $\cN= (0,2)$ supersymmetric sector which consists of a complex scalar $\phi$ and $\bar{\phi}$ 
together with a complex right-chirality fermion $\psi_+$ and $\bar{\psi}_+$. 
A natural generalization of the action \eqref{bosdef01}   is the following
\bsubeq \label{JTbar02comp}
\beqn 
S_{\alpha}&=&\int \rd^2 \sigma  \Big(\mathcal L_L+\mathcal L_R+\alpha\mathcal L_\text{def} \Big)  
~,\\
\mathcal L_L &=& \frac{\ri}{2} \bar \chi_- \p_{++} \chi_-  ~, \\ 
\mathcal L_R &=& \frac12 \p_{++} \bar\phi \p_{--}\phi - \frac{\ri}{2} \bar\psi_+ \p_{--} \psi_+  ~,\\ 
\mathcal L_\text{def} &=&  - \bar \chi_- \chi_- \Big(   \p_{++}\bar \phi \p_{++}\phi - \ri \bar\psi_+ \p_{++} \psi_+  \Big)   
~.
\eeqn
\esubeq
Let us check that this action describes a $J\bar T$ flow.

As for the $\cN=(0,1)$ case, 
the left-moving complex fermion has U(1) symmetry and the associated U(1) current is given by 
\be
G_{--}=\bar \chi_- \chi_-
~.
\ee
The $T_{++++}$ component of the  stress-energy tensor proves to be
\be\label{02Tpppp}
T_{++++}=-  \p_{++} \bar\phi \p_{++}\phi +\frac{\ri}{2} \bar\psi_+ \p_{++} \psi_+   +\frac{\ri}{2}  \psi_+ \p_{++}\bar \psi_+
~,
\ee
while the equation of motion for the fermion $\chi_-$ is
\be
\p_{++}\chi_-  =2\alpha\ri  \chi_-  T_{++++}
~.
\ee
Together with its complex conjugate, this  implies that 
the U(1) flavor current $G_{--}$ is chiral 
\be
\p_{++} G_{--}=0
~,
\ee
and that our action \eqref{JTbar02comp} arises from a chiral  $J\bar T$ deformation:
\be
\frac{\p  \mathcal L_\alpha}{\p  \alpha} =\mathcal L_\text{def}  =T_{++++} G_{--}
~.
\ee
However, it remains to show that the action is $\cN=(0,2) $ supersymmetric. 
For this we turn to superspace.

The fields $\phi$ and $\psi_+$ are going to describe a chiral $\cN=(0,2)$ multiplet while their complex conjugate 
fits in an anti-chiral one.
We can introduce the chiral and anti-chiral $\cN=(0,2)$ complex scalar superfields
\be
\Phi =\phi + \ri \sqrt2 \vartheta^+ \psi_+ +\ri \vartheta^+\bar \vartheta^+ \p_{++}\phi
~, \qquad
\bar\Phi =\bar \phi + \ri \sqrt2 \bar\vartheta^+ \bar\psi_+ -\ri \vartheta^+\bar \vartheta^+ \p_{++}\bar \phi
~,
\ee
satisfying the constraints
\be
\bar\cD_+ \Phi=\cD_+\bar\Phi=0
~.
\ee
We also introduce the $\cN=(0,2)$  complex Fermi-multiplet through the following superfields
\be
\Upsilon_ -=  \chi_- +  \vartheta^+ {\sf F}+\ri \vartheta^+ \bar \vartheta^+ \p_{++}\chi_-
~, \qquad
 \bar \Upsilon_ -= \bar  \chi_- +\bar\vartheta^+ \bar  {\sf F}- \ri  \vartheta^+ \bar \vartheta^+ \p_{++}\bar\chi_-
 ~.
\ee
These are also chiral and anti-chiral respectively
\be
\bar\cD_+ \Upsilon_ -  =\cD_+\bar \Upsilon_ -=0
~.
\ee
Note that the extra complex fields $ {\sf F}$ and $\bar{ {\sf F}}$ are necessary to close $\cN=(0,2)$ supersymmetry off-shell and will 
play the role of auxiliary fields, analogously to the $\cN=(0,1)$ case.
The natural ansatz  for the $J\bar T$-deformed action in superspace is then given by
\be\label{susy02JT}
\mathcal L_\alpha= \frac{1}{4} \int \rd\bar \vartheta^+ \rd\vartheta^+ \bar   \Upsilon_ -   \Upsilon_ -
+\frac{\ri}{4} \int \rd\bar \vartheta^+ \rd\vartheta^+ \bar\Phi \p_{--} \Phi  
-  \frac{ \alpha}{4}    \int \rd\bar \vartheta^+ \rd\vartheta^+ \bar\Upsilon_ -   \Upsilon_ - \cD_+ \Phi \bar \cD_+\bar\Phi
~.
\ee

We can compute the equation of motion of the chiral Fermi superfields that
gives
\be\label{02Dupsilon}
\cD_+\Upsilon_- 
= 2\ri\alpha   \Upsilon_- \cD_+ \Phi  \p_{++} \bar\Phi   
~,
~~~
\bar \cD_+ \bar  \Upsilon_- =  
 - 2\ri\alpha \bar   \Upsilon_- \bar  \cD_+ \bar \Phi  \p_{++} \Phi   
 ~.
\ee
The action \eqref{susy02JT} is invariant under the following symmetry: 
\be
\Upsilon_- \rightarrow  e^{\ri \Lambda} \Upsilon_-, \qquad  \bar \Upsilon_- \rightarrow  e^{-\ri \bar \Lambda} \bar \Upsilon_-
~,
\ee
with $\cD_+\bar \Lambda =\bar\cD_+  \Lambda=0$. 
Promoting this symmetry to a gauge symmetry, 
we can couple the Fermi multiplet to a real gauge prepotential superfield $V$ 
exactly in the same way as the well known 4D case:
$\bar\Upsilon_- e^V \Upsilon_-$. 
By expanding the resulting gauged action to leading order 
in $V$ and comparing to \eqref{02gaugecurrentcouple}, one can get the flavor current superfields
\be\label{02Gmm}
\cG_{--}=\bar \Upsilon_ -  \Upsilon_ -\Big(1-\alpha \cD_+\Phi \bar \cD_+\bar\Phi \Big)
~, \qquad \cG=0
~.
\ee
Noether  theorem guarantees that   $\cG_{--}$ is a conserved chiral current. Indeed, using 
the equation of  motion~\eqref{02Dupsilon},   one can verify that it holds
\be
\cD_+\cG_{--}=-\bar\Upsilon_- \cD_+  \Upsilon_ - \Big(1-\alpha \cD_+\Phi \bar \cD_+\bar\Phi \Big)
+2\ri \alpha \bar \Upsilon_ -  \Upsilon_ -  \cD_+\Phi   \p_{++} \bar\Phi =0
~.
\ee
The stress-tensor multiplet can also be straightforwardly computed. In particular, the $\cS_{++}$ superfield can be shown to be
\be
\cS_{++}= \frac12   \bar \cD_+\bar\Phi \cD_+ \Phi
~.
\ee
With this, one can compute $\cT_{++++}=\frac14 [\bar\cD_+, \cD_+] \cS_{++}$ 
and find that its lowest component  gives \eqref{02Tpppp}. 

The supersymmetric chiral $J\bar T$ primary operator, see eq.~\eqref{02JTbar-0} with $\cG\equiv 0$, 
 is thus given by 
\be
\mathcal O^{J\bar T}= \cS_{++}\cG_{--}= 
 \frac12   \bar \cD_+\bar\Phi \cD_+ \Phi
  \cdot \bar \Upsilon_ -  \Upsilon_ -\Big(1+\alpha \cD_+\Phi \bar \cD_+\bar\Phi \Big)
  = \frac12   \bar \cD_+\bar\Phi \cD_+ \Phi \bar \Upsilon_ -  \Upsilon_ -
  ~,
\ee
which is independent of deformation parameter $\alpha$.  
It is then obvious that  the action \eqref{susy02JT}  satisfies the following chiral supersymmetric  $J\bar T$ flow equation 
\be
\frac{\p\mathcal A_\alpha}{\p \alpha}
=-\frac14 \bar\Upsilon_ -   \Upsilon_ - \cD_+ \Phi \bar \cD_+\bar\Phi  
=\frac12 \cS_{++}\cG_{--}= \frac12\mathcal O^{J\bar T}
~.
\ee
Therefore, \eqref{susy02JT}  is indeed a manifestly off-shell $\cN=(0,2)$ supersymmetric $J\bar T$ deformed action. 

To analyze in more detail the action \eqref{susy02JT} and check its relation with \eqref{JTbar02comp}, 
we would like to expand the superspace action in components.
We find that  \eqref{susy02JT} after integrating the Grassmann variables gives
\beqn
\mathcal L_\alpha&=&\frac12 \p_{++} \bar\phi \p_{--}\phi  -  \frac{\ri}{2} \bar\psi_+ \p_{--} \psi_+ 
+\frac{\ri}{4} \bar\chi_- \p_{++}\chi_-+\frac{\ri}{4}  \chi_- \p_{++}\bar\chi_- -\frac 14  {\sf F}\bar  {\sf F} 
\non
\\&&
 -\frac{\alpha}{4} \Bigg[4 \chi_- \bar \chi_- \Big(-  \p_{++} \bar\phi \p_{++}\phi +\frac{\ri}{2} \bar\psi_+ \p_{++} \psi_+   
 +\frac{\ri}{2}  \psi_+ \p_{++}\bar \psi_+\Big)
 \non
\\&&\qquad
-2\bar\psi_+\psi_+ \Big( -\ri \chi_- \p_{++}\bar \chi_- -\ri \bar \chi_- \p_{++}\chi_-   + {\sf F}\bar  {\sf F}\Big)
\non\\&&\qquad
-2\sqrt 2 \bar\chi_- \bar \psi_+ \p_{++}\phi  {\sf F}
-2\sqrt 2   \psi_+  \chi_-  \p_{++} \bar \phi \bar  {\sf F} \Bigg]
~.
\eeqn
We stress that the previous Lagrangian leads to  an action which is $\cN=(0,2)$ supersymmetric off-shell. 
Solving the auxiliary field equations of motion gives
\be
\bar  {\sf F}=2\sqrt 2\alpha \bar\chi_- \bar \psi_+ \p_{++}\phi
~, \qquad   {\sf F}=2\sqrt 2  \alpha \psi_+  \chi_-  \p_{++} \bar \phi 
~.
\ee
By using this result 
the action turns into
\beqn\label{02compoentFull}
S_\alpha&=&\int \rd^2 \sigma \Bigg[ \, \frac12 \p_{++} \bar\phi \p_{--}\phi  -  \frac{\ri}{2} \bar\psi_+ \p_{--} \psi_+ 
+\frac{\ri}{4}\Big(   \bar\chi_- \p_{++}\chi_-+   \chi_- \p_{++}\bar\chi_- \Big) \Big(1-2\alpha \bar \psi_+\psi_+\Big)
\qquad
\non\\&&\qquad\qquad
 -\alpha   \bar \chi_-  \chi_- \Big(  \p_{++} \bar\phi \p_{++}\phi - \frac{\ri}{2} \bar\psi_+ \p_{++} \psi_+ 
   -\frac{\ri}{2}  \psi_+ \p_{++}\bar \psi_+\Big)
\non\\&&\qquad\qquad
-2 \alpha^2\chi_- \bar \chi_-   \bar \psi_+\psi_+   \p_{++} \bar\phi \p_{++}\phi  
\Bigg]
~.
\eeqn
Compared to  the original component action \eqref{JTbar02comp}, we see that there are two extra pieces: one is the $\alpha^2$ 
term, and the other one multiplies the kinetic term of the left fermions.   
As we will see, these extra terms can be redefined away.

Note that the flavor supercurrent is given by \eqref{02Gmm} and its lowest component gives the conventional flavour current
\be\label{02Gmmfull}
G_{--} = \bar \chi_ -\chi_- (1-2\alpha \bar \psi_+\psi_+)
~.
\ee
Following our previous superfield approach, this current is  a chiral conserved current $\p_{++}G_{--}=0$. 

To see the role of this current, we can   rewrite the action  \eqref{02compoentFull}   in the   following form: 
\bea\label{02compoentFullanother}  
S_\alpha&=&\int \rd^2 \sigma \Bigg[ \, \frac12 \p_{++} \bar\phi \p_{--}\phi   -  \frac{\ri}{2} \bar\psi_+ \p_{--} \psi_+ 
+\frac{\ri}{4}\Big(   \bar\chi_- \p_{++}\chi_-+   \chi_- \p_{++}\bar\chi_-  \Big) \Big(1-2\alpha \bar \psi_+\psi_+\Big)
\non\\&&\qquad~~~
 -\alpha   \bar \chi_-  \chi_- \Big(1-2\alpha \bar \psi_+\psi_+\Big) \Big(  \p_{++} \bar\phi \p_{++}\phi 
 - \frac{\ri}{2} \bar\psi_+ \p_{++} \psi_+   -\frac{\ri}{2}  \psi_+ \p_{++}\bar \psi_+\Big)
\Bigg]
~.~~~~~~~~~
\eea
The first line  can be thought as  the undeformed action where the left-moving fermion still has a U(1) symmetry. 
The associated current   is  exactly given by \eqref{02Gmmfull}.
 And the second line is then just the $J\bar T$ deformation with a modified current \eqref{02Gmmfull}. 

The action \eqref{02compoentFullanother}   can also be obtained from \eqref{JTbar02comp} through a field redefinition: 
\be
\chi_- \to  \chi_- \Big(1- \alpha \bar \psi_+\psi_+\Big)~, \qquad
\bar\chi_- \to \bar \chi_- \Big(1- \alpha \bar \psi_+\psi_+\Big)~. 
\ee
To conclude, the supersymmetric $J\bar T$ deformation in  \eqref{susy02JT}, \eqref{02compoentFullanother}  
 and  the conventional  $J\bar T$ deformation in \eqref{JTbar02comp}  coincide up to field redefinitions. This  implies  that  these actions are the same  on-shell, as  expected from the general equivalence
 of the manifestly supersymmetric $J\bar T$ deformation and the one given by the operator $O^{J\bar{T}}_{++}$, 
 eq.~\eqref{nonsusyJTbar}.

\subsection{$\cN=   (0,2)$   $T\bar J$}

In this subsection we shortly present a
model for an $\cN=   (0,2)$   $T\bar J$ deformation which extends the bosonic and $\cN=(0,1)$ cases presented in section
\ref{TJbar01}. To make the presentation more concise and manifestly supersymmetric, we will work directly in superspace.

We are going to show that the following model
\be\label{susy02TJ}
S_\lambda=\int \rd^2 \sigma \rd\bar \vartheta^+ \rd\vartheta^+ \mathcal A_\lambda
= \frac{\ri}{4} \int \rd^2 \sigma \rd\bar \vartheta^+ \rd\vartheta^+\;  \bar\Phi \p_{--} \Phi   F(\lambda \p_{--}\bar\Phi)
~,
 \ee
 is a $T\bar J$ flow and, in particular,
 a $\cN=   (0,2)$   extension of  the action \eqref{TJdefactionboson}.
 
By  considering the variation with respect to the $\cN=(0,2)$ chiral superfield $ \Phi$, we get 
the following  equation of motion
 \be
 \p_{--} \bar\cD_+ (\bar\Phi F)=0
 ~.
 \ee
Compared with \eqref{02flavorcurrent0}, we can naturally identify the following chiral   U(1) current 
\footnote{Note that this current is now complex, so only the first equation in  \eqref{02flavorcurrent0} is satisfied. 
We will comment on it later. }
\be
\cG  =-\ri\gamma \bar\Phi F
~,\qquad
\cG_{--}=0
~,
\ee
where $\gamma $ is an arbitrary normalization constant introduced for convenience. 
In the undeformed limit $F=1$,
 this is indeed the U(1) current associated with the shift symmetry of the superfield $\F$. 

It is also straightforward to compute the $\cT_{----}$ component of the stress-tensor multiplet which can be shown to be
\be
\cT_{----}= - \p_{--}  \bar\Phi \p_{--} \Phi   F 
~.
\ee

By requiring that the $\mathcal A_\lambda$ superfield Lagrangian satisfies the chiral $T\bar J$ flow equation~\eqref{TJbar02R}
\be
\frac{\p\mathcal A_\lambda}{\p\lambda} =\cO_{----}^{T\bar J}=\cT_{----}\cG
~,
\ee
we obtain the following condition for the function $F$
\be
\frac{\ri}{4}  \bar\Phi \p_{--}\Phi   \p _{--}\bar\Phi  F'=\ri  \gamma\bar\Phi  \p_{--}  \bar\Phi \p_{--}\Phi   F^2~ \qquad\Longrightarrow\qquad F'=4 \gamma F^2
~.
\ee
By imposing $F(0)=1$,   the solution is 
\footnote{One can then choose the normalization $\gamma=\frac{1}{16}$ such that the solution agrees with \eqref{solFmonica}.}
\be
F(x)=\frac{1}{1-4 \gamma x}
~.
\ee
 
Therefore, we have shown that the action $S_\l$ in \eqref{susy02TJ} 
satisfies a chiral $T\bar J$ flow. 
However, it is clear that the action is actually a little pathological because it is not real due the complex 
chiral current $\cG(\z)$ and its descendant
$G_{++}(\s):= -\frac12 [\cD_+, \bar \cD_+] \cG(\z)|_{\vartheta=0}$.
This   does not change or spoil the basic properties of the $T\bar J$ deformation.
However, it would be interesting and important  to see whether one could modify the action to get a theory arising from a chiral 
 $T\bar J$ deformation with \emph{real} U(1) current. We leave this for future analysis.

\section{Conclusion and outlook}  \label{conclusion}

In this paper we have analyzed
$J\bar T /T \bar J$-deformations 
for theories possessing  $\cN=(0,1),\,(1,1) $ and $(0,2)$ supersymmetry.  
We have first discussed
the conservation equations of the stress-tensor multiplets and 
flavor current multiplets. 
Based on those multiplets,
we have then  constructed the   $  J\bar T /T \bar J$ supersymmetric primary operators. 
We have further shown that their descendants are equivalent to the conventional $\ J\bar T /T \bar J$ operators 
up to conservation equations
and total derivatives. Several examples of Lagrangians arising from the  chiral $  J\bar T /T \bar J$ deformation of free 
supersymmetric theories were  also presented.  

To construct the $  J\bar T /T \bar J$ operator, a  conserved U(1) current is needed.  In this paper we have been focusing
 exclusively on a flavor  U(1) current that does not belong to the stress-tensor multiplet. 
 However, in some supersymmetric theories, there is  also an  $R$-symmetry which 
can give rise to a  U(1) $R$-current. A natural question is: can we construct the $  J\bar T /T \bar J$ operators 
out of the stress-energy tensor 
and the U(1) $R$-current? In our $\cN=(0,2) $ case, the $\cR$-multiplet  is given in \eqref{02Rmulti} and it contains both the 
stress-energy
tensor $T_{++\pm\pm } $ and $R$-current $j_{\pm\pm}$ which enables one to construct the conventional   $  J\bar T /T \bar J$ 
operator. However, a supersymmetric primary 
built out of the  $\cR$-multiplet seems to evade  the constructions in this paper. 
It would be  interesting  to investigate in detail
the underlying reasons of the failure/success of 
these $R$-symmetry deformations  and analyze them also 
in theories with more supersymmetries, say $\cN=(2,2)$. 
We leave this problem for the future.

In our paper, we have only considered Lagrangians arising from the  chiral $  J\bar T /T \bar J$ deformations of free theories because   
these deformed models are simple and argued to be solvable. 
Starting from the relativistic free theory, we indeed find several 
simple theories arising from chiral $  J\bar T$  and $T \bar J$ deformations with  $\cN=(0,1)$ 
supersymmetry and a chiral $ J \bar T$ deformation  with  
$\cN=(0,2)$ supersymmetry. However, in the $\cN=(1,1)$ case, we did not find a simple realization of chiral 
$  J\bar T /T \bar J$ deformations.%
\footnote{For example, the   $J\bar T $ construction in subsection~\ref{01JTbarExample} is not obvious  because the $\cN=(1,1)$ 
supersymmetric generalization of the left moving sector in \eqref{01JTbarLeft} requires the  embedding of the \emph{complex} 
fermion $\chi_-$ into a superfield which necessarily introduces also many other fields. 
For the naive  $\cN=(1,1)$  supersymmetric generalization of  $T\bar J $ construction in 
\eqref{TJdefactionboson}, the EoMs contain many types of derivatives $\cD_\pm , \p_{\pm\pm}$ and thus fails to guarantee the 
chiral conservation of the U(1)  current $\cD_+\cG_-=0 $ or $\cD_-\cG_+=0$
in a simple way.  }
 It would be interesting to see whether this type of $\cN=(1,1)$ chiral 
 $  J\bar T /T \bar J$ deformations can be realized in a more complicated or  broad 
 class of theories. For example, since $  J\bar T /T \bar J$ deformations break  Lorentz invariance, one can naturally start with a 
 non-Lorentz-invariant but supersymmetric theory   and see whether it admits a chiral $  J\bar T /T \bar J$ deformation with some 
 amount of supersymmetry.
In  the   $\cN=(0,2)$ case, as an example, 
we have presented a chiral  $T\bar J$ deformed Lagrangian with \emph{complex} current. It remains to 
see  how to construct a    \emph{real}  chiral  $T\bar J$ deformed theory with   $\cN=(0,2)$ supersymmetry.

 Another question is the symmetry enhancement. As argued in~\cite{Guica:2017lia,Bzowski:2018pcy},  the $  J\bar T /T \bar J$  
 deformation breaks the  original  two-dimensional  conformal group $SL(2,  \mathbb R ) \times  SL(2,  \mathbb R )$   down to     the 
 $SL(2,  \mathbb R ) \times  U(1)$ subgroup as the global symmetry of the deformed theory, but these symmetries would be 
 enhanced to the infinite-dimensional  Virasoro  $\times $ Virasoro. Now with supersymmetries, it is natural to expect that the 
 enhancement is given by a super-Virasoro  $\times$ super-Virasoro symmetry.%
 \footnote{Besides,
 there is also a chiral  $U(1)_J$ symmetry generated by the current; this symmetry is now expected to  enhance to  
 super-Kac-Moody.  }
 
Last but not least, as shown in appendix~\ref{SZgeneralization}, 
all the operators associated to $T\bar T$ and $  J\bar T /T \bar J$  deformations
fit into 
a general pattern which generalizes the Smirnov-Zamolodchikov type of composite operators. 
In appendix~\ref{SZgeneralization}, 
we have also shown that under certain 
assumptions, the generalized composite operator is invariant under improvement transformations.
The original Smirnov-Zamolodchikov type   composite operators are proved  to be well-defined  at the quantum level. For our 
generalization, this quantum definedness has also been shown to hold in several examples explicitly. It is thus reasonable to 
speculate that our generalized Smirnov-Zamolodchikov composite operators are 
also well-defined  at the quantum level in general.  The  proof of this statement
and its implications will be  an interesting  and important future research problem.


\section*{Acknowledgements}
 
We are grateful to Alessandro Sfondrini for collaboration at early stages of  this project
and to Christian Ferko for comments on the manuscript. 
G.~T.-M.  would also like to thank the organizers and participants of the workshop on 
\emph{``$T \bar{T}$ and Other Solvable Deformations of Quantum Field Theories''} 
for providing a stimulating atmosphere, and the Simons Center for Geometry and Physics for hospitality and partial support.    
  We are also grateful for the support and vively atmosphere during the workshop
\emph{``New frontiers of integrable deformations''}
in Villa Garbald, Castasegna.  
H.~J. is supported by the Swiss National Science Foundation.
The work of G.~T.-M. was supported by the Albert Einstein Center for Fundamental Physics, University of Bern,
by the Australian Research Council (ARC) Future Fellowship FT180100353,
and by the Capacity Building Package of the University of Queensland.


 \appendix

 \section{Deriving the  conservation laws of   flavor current multiplets} \label{DeriveFlavorCurrent}
 
 In this appendix we derive the various flavor current multiplets described in section \ref{flavorCurrentMultiplet}.
 The derivation is conceptually the same for all the types of supersymmetries.
 As a first step we describe a supersymmetric abelian vector multiplet and its  gauge transformation
rules. 
Then we couple the gauge multiplet to a corresponding flavor current multiplet
and impose the gauge invariance of such coupling.
As a result we obtain the conservation equations of the supersymmetric flavor current multiplets.

 \subsection{$\mathcal N=(0,1)$} \label{DeriveFlavorCurrent01}

 By looking for example at \cite{Gates:1986ez}, pages 5-6, we see that an $\cN=(1,0)$ abelian vector multiplet is 
described by a gauge connection $\G_A$ and gauge covariant derivatives
\bea\label{covD01}
\nabla_A=\cD_A-\ri\G_A
~,
\eea
satisfying the following algebra
\bsubeq\label{gauge01}
\bea
&\{\nabla_+,\nabla_+\}=-2\ri\nabla_{++}~,
\label{gauge01-1}
\\
&[\nabla_+,\nabla_{--}]=\ri \cW_-
~,\qquad
[\nabla_+,\nabla_{++}]=0
~,
\label{gauge01-3/2}
\\
&[\nabla_{++},\nabla_{--}]=-\nabla_+\cW_-
~.
\label{gauge01-2}
\eea
\esubeq
Here the superfield $\cW_-(\z)$ is an unconstrained real spinorial field strength.
The previous algebra correctly satisfies the super-Jacobi identities and in fact
it is interesting to note that the form of the commutator $[\nabla_{++},\nabla_{--}]$
is fixed by the   Bianchi identities
\bea
[\nabla_{++},\nabla_{--}]&=&\frac{\ri}{2}[\{\nabla_+,\nabla_+\},\nabla_{--}]=\ri\{\nabla_+,[\nabla_+,\nabla_{--}]\}
=-\{\nabla_+,\cW_-\}
=-\nabla_+\cW_- 
~.~~~~~~~~~
\eea
 The anti-commutator \eqref{gauge01-1} implies that $\G_{++}$ can be solved  in terms of $\G_+$%
\footnote{Note that we are considering an Abelian gauge symmetry, so that the connections 
(anti-)commute. }
\bea
\G_{++}=\ri\cD_+\G_+
~,
\label{Gamma++D+Gamma+}
\eea
while $\G_+$ and $\G_{--}$ remain independent and unconstrained gauge connections. 
The $\cN=(0,1)$ superfields $(\G_+(\z),\,\G_{--}(\z))$ then play exactly the same role of unconstrained component gauge connection 
fields, $(A_{++}(\s),\,A_{--}(\s))$, gauging an Abelian  symmetry in the standard two-dimensional Minkowski space-time.
The first  equation in \eqref{gauge01-3/2} can be   used to express $\cW_-(\z)$   in terms of the unconstrained connections
\bea
\cW_-=-\cD_+\G_{--}+\p_{--}\G_+
~.
\label{W-DG}
\eea
All the other constraints associated with the algebra \eqref{gauge01} are then identically satisfied
once \eqref{Gamma++D+Gamma+} and \eqref{W-DG} are imposed.

Note that the gauge transformations of $\G_{--}$ and $\G_+$ are 
\bea\label{gaguetsf01}
\d_G\G_{--}=\ri\p_{--}\t
~,~~~~~~
\d_G\G_{+}=\ri\cD_+\t
~,
\eea
with $\t(\z)$ an unconstrained real gauge superfield parameter. It is easy to see that 
\bea
\d_G\cW_-=0
~,
\eea
so the field strength is gauge invariant as expected.

In components, the multiplet of connections reads
\be 
\Gamma_+(\z) 
= 
 \chi_+(\s) +\vartheta^+ A_{++}(\s)
~,   \qquad
\Gamma_{--}(\z)
= 
\ri A_{--}(\s) 
-\vartheta^+ \lambda_-(\s) 
~,
\ee 
and the field strength is
\be\label{01Wm}
\cW_-(\z)
 = 
\lambda_- (\s)
+\p_{--}\chi_+(\s) 
+\vartheta^+ \Big(\p_{--} A_{++}(\s) -\p_{++} A_{--}(\s) \Big) 
~.
\ee
Then under the gauge transformation \eqref{gaguetsf01} with 
$\t(\z) = \phi(\s)+\vartheta^+ \psi_+(\s) $, the component fields transform as
\be
\delta_G \chi_+ = \ri \psi_+
~, \qquad 
\delta_G \lambda_- = -\ri \p_{--}\psi_+
~, \qquad  
\delta_G A_{++} =   \p_{--}\phi
~, \qquad \delta_G A_{--} =   \p_{++}\phi
~.
\ee
These transformations obviously leave the components of the field strength \eqref{01Wm} invariant. 
Furthermore, they imply that $\chi_+$ 
is  pure gauge and can be set to zero. 
Then the two independent components of the field strength multiplet are the gaugino $\l(\s)$ and the field strength $F(\s)$:
\be
\l_-(\s)=\cW_-(\z)|_{\vartheta=0}
~, \qquad
F(\s)=\nabla_+\cW_-(\z)|_{\vartheta=0}=\p_{--} A_{++}(\s) -\p_{++} A_{--}(\s)
~.
\ee
Note that  $F$ is a pseudo-scalar field that 
 arises from the Hodge dual of the field strength $F_{ab}=\p_{[a}A_{b]}$.

Now that we have reviewed the structure of an $\cN=(0,1)$ vector multiplet,
we can derive the multiplet of currents for an Abelian symmetry.
Consider a U(1) invariant action $S$ for a matter system. 
If we couple it to a background U(1) gauge  multiplet
described by the independent superfields $(\G_+,\G_{--})$, 
at first order in the   gauge connections it holds
\begin{equation}\label{01gaugecopuling}
S=-\ri\int\rd^2\s\,\rd\vartheta^+\Big{[} 
\cG_+\G_{--}
+\ri\cG_{--}\G_{+}
\Big{]}
\,.
\end{equation}
Assuming that the equations of motion for the matter multiplets are satisfied,
the variation of the action under arbitrary local U(1) transformations~\eqref{gaguetsf01}, 
after some integrations by parts, takes the form
\bea
\d_G S=\ri\int\rd^2\s\,\rd\vartheta^+\,\t\big(
\p_{--}\cG_+
+\ri\cD_+\cG_{--}
\big)
~.~~~
\eea
Imposing that the action is invariant $\d_G S=0$ then leads to the following supercurrent conservation equations
for  a U(1) symmetry:
\bea
\cD_+\cG_{--}=\ri\p_{--}\cG_+
~.
\eea
It is simple to see that the previous conservation equation implies
\bea
\ri\p_{--}\cD_+\cG_+=\cD_+\cD_+\cG_{--}=-\ri\p_{++}\cG_{--}
~.
\eea
Thus by defining 
\bea
\cG_{++}:=\cD_+\cG_+
\label{G++D+G+}
~,
\eea
one gets the conservation equation for a U(1)  flavor current
\bea
\p_{++}\cG_{--}=-\p_{--}\cG_{++}
~.
\label{01vectorsuperfieldsconserv}
\eea
Note that by construction, due to \eqref{G++D+G+}, it also holds
\bea
\cD_+\cG_{++}=-\ri\p_{++}\cG_+
~.
\eea

In components, the superfields of the U(1) flavor current multiplet are given by
\be
\cG_+(\z) =g_+(\s) 
+\vartheta^+ G_{++}(\s)
~, \qquad 
\cG_{--}(\z) =G_{--}(\s) 
+\ri \vartheta^+ \p_{--} g_+(\s)
~.
\ee
Due to eq.~\eqref{01vectorsuperfieldsconserv}, 
 $G_{\pm\pm}$ satisfy the ordinary vector conservation equation
\be
\p_{--}G_{++}+\p_{++}G_{--}=0
~.
\ee

 \subsection{$\mathcal N=(1,1)$ } \label{DeriveFlavorCurrent11}
 
 The Abelian current multiplet with $\cN=(1,1)$  supersymmetry can be derived in a similar fashion as that in the $\cN=(0,1)$ case. 
 In practice, we can appropriately combine the two copies of $\cN=(0,1)$  and $\cN=(1,0)$ currents
  that arise from parity transformations of one to the other. A description of the off-shell vector multiplet for $\cN=(1,1)$ 
  can be found in \cite{Ferrara:1975nf}.
  
 The superspace Abelian gauge covariant derivatives are given in terms of connections $\G_A(\z)$ by
 \be
 \nabla_A= \cD_A-\ri\Gamma_A
 ~,
 \ee
 where the flat spinor derivatives are given in \eqref{11DandQ}. 
To describe an irreducible vector multiplet,
the covariant derivatives are constrained to satisfy the following algebra
\bsubeq\label{gauge11}
\bea
& \{\nabla_+,\nabla_+\}=-2\ri\nabla_{++}
~,  \quad 
 \{\nabla_-,\nabla_-\} =-2\ri\nabla_{--}
~,   \quad
\{\nabla_+,\nabla_-\}=-\ri \cW
 ~,
 \label{gauge11-1}
\\
&  [\nabla_+,\nabla_{--}]=-\nabla_-\cW
~,  \quad  
 [\nabla_-,\nabla_{++}] =-\nabla_+\cW
 ~,   \quad
  [\nabla_-,\nabla_{--}]=[\nabla_+,\nabla_{++}]=0
  ~,~~~~~~~~~
   \label{gauge11-3/2}
\\
&  [\nabla_{++},\nabla_{--}]=-\ri\nabla_+\nabla_-\cW
  ~.
   \label{gauge11-2}
\eea
\esubeq

We would like to describe the previous algebra completely in terms of independent connections.
By analyzing the first two anti-commutators in
\eqref{gauge11-1} we can express   the vector connections $\G_{\pm\pm}$ in terms of the spinor ones $\G_{\pm}$
as
\be
\Gamma_{++}=\ri \cD_+\Gamma_+
~, \qquad \Gamma_{--}=\ri \cD_-\Gamma_-
~.
\ee
Moreover, from the third anti-commutator in
\eqref{gauge11-1} we obtain the expression of the scalar superfield strength $\cW(\z)$ in terms of the independent connections
$\G_{\pm}(\z)$
\be
\cW=\cD_+ \Gamma_-+\cD_-\Gamma_+ 
~.
\label{W11}
\ee
With these relations holding, it is easy to verify that 
the rest of the algebra is completely determined in terms of the unconstrained connection superfields $\G_+$ and $\G_-$.

The gauge transformation is given by 
\be\label{gaugetsf11}
\delta_G\Gamma_+=\ri \cD_+\tau
~, \qquad  \delta_G\Gamma_-=\ri \cD_-\tau
~.
\ee
It leaves the field strength  invariant $\delta_G \cW=0$.

Note that in components, the previous Abelian vector multiplet is reduced in the following way.
The connections are
\bsubeq\label{02Gpm}
\bea
\Gamma_+(\z) &=&\chi_+(\s)+\vartheta^+ A_{++}(\s)
+\vartheta^-B_{-+}(\s)
+\ri \vartheta^+ \vartheta^- \eta_+(\s)~,
\\
\Gamma_-(\z) &=&\chi_-(\s) 
+\vartheta^- A_{--}(\s)
+\vartheta^+B_{+-}(\s)
+\ri \theta^- \theta^+ \eta_-(\s)
~.
\eea
\esubeq
The field strength $\cW$ is consequently given by 
\bea
\cW(\z)
&=&
B_{-+}(\s)+B_{+-}(\s)
 -\ri \vartheta^+\big(\eta_+(\s)+\p_{++}\chi_-(\s)\big)
  -\ri \vartheta^-\big(\eta_-(\s)+\p_{--}\chi_+(\s)\big)
  \non\\
  &&
-\ri\vartheta^+\vartheta^- \big(\p_{++}A_{--}(\s) -\p_{--}A_{++}(\s)\big)
~.
\eea

Under the gauge transformation \eqref{gaugetsf11} with  gauge parameter  
\be
\tau(\z)=\phi(\s)
+\ri\vartheta^+ \psi_+(\s) +\ri \vartheta^- \psi_-(\s)  +\ri \vartheta^+ \vartheta^- C (\s)
~.
\ee
the connections~\eqref{02Gpm} transform as
\bsubeq
\bea
\delta_G \Gamma_+&=&\ri \cD_+ \tau =-\psi_++\vartheta^+\p_{++} \phi -\vartheta ^- C +\ri \vartheta^+\vartheta^-\p_{++}\psi_-
~,
\\
\delta_G \Gamma_-&=&\ri \cD_- \tau =-\psi_- +\vartheta^-\p_{--} \phi +\vartheta ^+ C -\ri \vartheta^+\vartheta^-\p_{--}\psi_+
~.
\eea
\esubeq
One can check that under this gauge transformation, the components of $\cW$ are indeed invariant. 
We can choose a WZ gauge such that  $\chi_+=\chi_-=0$, then 
\be
\cW=B-\ri \vartheta^+ \eta_+ -\ri \vartheta^- \eta_- 
-\ri\vartheta^+\vartheta^-F
~,
\ee
where 
\be
B=B_{-+}+B_{+-}~ , \qquad F=\p_{++}A_{--} -\p_{--}A_{++}
~.
\ee
Then 
the physical degrees of freedoms include two real gaugni $\eta_\pm$ and one real scalar $B$ as well as one pseudo-real scalar 
$F$ \cite{Ferrara:1975nf}.

As before for the $\cN=(0,1)$ case, we can couple the vector multiplet to the Abelian current superfields $\cG_\pm(\z)$:
\be
S=\int \rd^2 \sigma \rd \vartheta^+ \rd\vartheta^- \Big( \Gamma_- \cG_+ -\Gamma_+ \cG_-   \Big)
~.
\ee
 Under the gauge transformation~\eqref{gaugetsf11}, the action transforms as 
 \be
 \delta_G S=-\ri \int \rd^2 \sigma \rd \vartheta^+\rd\vartheta^-  \; \tau \Big(   \cD_+ \cG_-  -   \cD_- \cG_+  \Big)
 ~.
 \ee
By imposing gauge invariance, we obtain the conservation equation for the U(1) current
 \be\label{U1sym}
   \cD_+ \cG_-  -   \cD_- \cG_+=0
   ~.
 \ee
 We can define the descendant superfields
 \be
  \cG_{++}=  \cD_+ \cG_+
  ~, \qquad     \cG_{--}=  \cD_- \cG_-
  ~.
 \ee
Then acting with $\cD_+\cD_-$ on  both sides of equation  \eqref{U1sym} yields 
\be\label{11Gppmm}
\p_{--} \cG_{++}+\p_{++}\cG_{--}=0
~.
\ee

In components, the U(1) current multiplet reads
\bsubeq
\beqn
\cG_+&=&g_+ +\vartheta^+ G_{++} +\vartheta^- p + \ri \vartheta^+ \vartheta^- \p_{++} g_-
~, \\
\cG_-&=&g_- +\vartheta^- G_{--} +\vartheta^+ p- \ri \vartheta^+ \vartheta^- \p_{--} g_+
~.
\eeqn
\esubeq
The lowest component of \eqref{11Gppmm} is just the conventional U(1) vector current conservation equation
\be
\p_{--}  G_{++}+\p_{++} G_{--}=0
~.
\ee

 \subsection{$\mathcal N=(0,2)$} \label{DeriveFlavorCurrent02}

In this section, we will first review the gauge multiplet with $\cN=(0,2) $ supersymmetry following~\cite{Brooks:1986gd}. 
After that, by following the same standard approach used above for the $\cN=(0,1)$ and $\cN=(1,1)$ cases,
we will derive the current multiplet for $\cN=(0,2) $  supersymmetric theories. 

The Abelian vector multiplet can be constructed by introducing the gauge covariant derivatives:
\be
\nabla_+ =\cD_+-\ri \Gamma_+
~, \quad 
\bar \nabla_+ =\bar\cD_+-\ri \bar\Gamma_+
~, \quad
 \nabla_{\pm\pm} =\p_{\pm\pm} -\ri \Gamma_{\pm\pm}
 ~, 
\ee
where the spinor covariant derivatives were introduced in \eqref{02DandQ}.
Note also the conjugation properties  
 $\bar\cD_+=-\big(\cD_+\big)^\dagger, \bar\Gamma_+=-\big(\Gamma_+\big)^\dagger, \bar\nabla_+=-\big(\nabla_+\big)^\dagger$.

An irreducible vector multiplet is obtained by
imposing the following constraints on the algebra:%
\footnote{Note the conjugation properties: $\big( \cF \big)^\dagger=\cF, \big( \cW_- \big)^\dagger=-\bar \cW_-$.
Note also that the field strengths $\cW_-$ and $\bar{\cW}_-$ should not be confused with the trace currents
of the $\cN=(0,2)$ stress-tensor multiplet used in section \eqref{02stressTensor}.}
\bsubeq
\bea\label{02DpDp}
&\{\nabla_+,   \nabla_+\}= \{\bar \nabla_+, \bar \nabla_+\}=[\nabla_+,   \nabla_{++}]=[\bar\nabla_+,   \nabla_{++}]=0
~, \quad 
 \{ \nabla_+, \bar  \nabla_+\}=2\ri \nabla_{++}
~,~~~~~~~~~
\\  \label{02Wstrength}
&  [\nabla_+,  \nabla_{--}]=-\ri \bar \cW_-
~,  \qquad 
 [\bar\nabla_+,  \nabla_{--}]=-\ri   \cW_-
~, \quad 
 [\nabla_{++}, \nabla_{--}] = -\ri \cF
 ~,~~~~~~~~~
\eea
\esubeq
where the superfield strengths satisfy the following Bianchi identities 
\bsubeq\label{BI02}
\bea
&\nabla_+ \bar \cW_-=\bar \nabla_+ \cW_-=0
~, \quad  
\nabla_+ \cW_- +\bar  \nabla_+ \bar \cW_-=2\ri \cF
~, 
\\
&\nabla_+ \cF=\nabla_{++} \bar \cW_-
~,\quad \bar \nabla_+ \cF=\nabla_{++} \cW_-
~.
\eea
\esubeq
These imply
\be
\nabla_+ \cW_-= \cR+\ri  \cF
~, \qquad   
\bar \nabla_+  \bar \cW_-= -\cR+\ri \cF
~, \qquad 
\big( \cR\big)^\dagger=\cR
~.
\ee

We are interested in the Abelian gauge theory.  
It is easy to show that the vanishing of the first two anti-commutators in \eqref{02DpDp} gives  
$\cD_+\Gamma_+=\bar\cD_+\bar\Gamma_+=0$. 
Since it holds $\cD_+^2=\bar\cD_+^2=0$, 
we can  rewrite the spinor connections  in terms of the real unconstrained prepotential $V$ as
\bsubeq
\beqn
\Gamma_+&=&\ri e^{-V} \cD_+ e^V=\ri \cD_+ V
~,
\\
\bar\Gamma_+&=&\ri e^{ V} \bar\cD_+ e^{-V}=-\ri \bar \cD_+ V
~.
\eeqn
\esubeq
Moreover,  the last anti-commutator in \eqref{02DpDp} expresses the vector connection $\G_{++}$ in terms of the spinor ones:
\be
\Gamma_{++} =-\frac{\ri}{2} (\cD_+ \bar \Gamma_++\bar \cD_+ \Gamma_+)
~.
\ee
From \eqref{02Wstrength}, we can obtain the following expressions for the  superfield strengths  
\bsubeq\beqn
\bar \cW_- &=&-\p_{--} \Gamma_++\cD_+ \Gamma_{--} 
~, \\
  \cW_- &=&-\p_{--} \bar\Gamma_++ \bar\cD_+ \Gamma_{--} 
  ~, \\
\cF &=& \p_{++} \Gamma_{--} - \p_{--}    \Gamma_{++} =  \p_{++} \Gamma_{--} 
   +\frac{\ri}{2} \p_{--}   (\cD_+ \bar \Gamma_++\bar \cD_+ \Gamma_+)  
   ~,
\eeqn
\esubeq
which satisfy the Bianchi identities \eqref{BI02}. As a result, the unconstrained gauge fields for the $\cN=(0,2)$ vector multiplet
are the real prepotential $V$ and the  connection $\G_{--}$.

The gauge transformation of the prepotential $V$ is given by 
\be
\delta_G V=\ri(\Lambda-\bar \Lambda )
~,
\ee
where $\Lambda$ and $\bar\Lambda$ are chiral and anti-chiral, respectively:
\be
\cD_+ \bar \Lambda=\bar \cD_+ \Lambda=0
~.
\ee
As a consequence, the connections transform as  
\bsubeq\beqn
\delta_G \Gamma_+ & = &-  \cD_+ \Lambda
~,\\ 
\delta_G  \bar \Gamma_+ & = &  -\bar \cD_+  \bar \Lambda
~,\\ 
\delta_G \Gamma_{--} & = &-\p_{--}( \Lambda + \bar \Lambda)
~,\\ 
\delta_G \Gamma_{++} & = &-\p_{++}( \Lambda + \bar \Lambda) 
~.
\eeqn
\esubeq
It is easy to  verify that these gauge transformations leave  the field strengths invariant:
\be
\delta_G \cW_-=\delta_G \bar \cW_-=\delta_G \cF=0
~.
\ee

By using the $\cN=(0,2)$ Abelian vector multiplet described above,  we can now derive the U(1) current multiplet. 
We proceed by coupling the unconstrained gauge potentials $\Gamma_{--}$ and $V$
to an Abelian  current multiplet in the following way
\be\label{02gaugecurrentcouple}
S=\int \rd^2 \sigma \rd \bar \vartheta^+ \rd\vartheta^+ \Big( \Gamma_{--}  \cG  +V \cG_{--} \Big)
~,
\ee
where $\cG_{--}$ and $  \cG$ are real superfields. 
Under a gauge transformation, the previous action transforms as 
\beqn
\delta_G S
&=&\int \rd^2 \sigma \rd \bar \vartheta^+ \rd\vartheta^+  
\Big(   \Lambda ( \p_\mm \cG +\ri \cG_{--})    +\bar \Lambda( \p_\mm \cG -\ri \cG_{--})    \Big)
~.
\eeqn
Note that $\Lambda$ is a chiral superfield while $\bar \Lambda$ is an anti-chiral superfield. 
Hence, the gauge invariance leads to the following two conservation equations
\be
\bar \cD_+(    \cG_{--}  -\ri  \p_\mm \cG )   =0
~, \qquad   \cD_+(    \cG_{--}  +\ri  \p_\mm \cG )  =0
~,
\ee
that are conjugate to each other. If we define the descendant superfield
\be
\cG_{++}= -\frac12 [\cD_+, \bar \cD_+] \cG
~,
\ee
then it is straightforward to prove that the following vector conservation eqution
\be\label{02Gppmm}
\p_{++} \cG_{--}+\p_{--}\cG_{++}=0
\ee
holds.

In components, the current multiplet is given by
\bsubeq
\beqn
\cG(\z)&=&g(\s)
+\ri \vartheta^+ p_+(\s)
 +\ri\bar\vartheta^+ \bar p_+(\s) 
 +\vartheta^+\bar\vartheta^+ G_{++}(\s)
 ~,
\\
\cG_{--}(\z)
&=&
G_{--}(\s)+\vartheta^+ \p_{--} p_+(\s)
 -\bar\vartheta^+ \p_{--}\bar p_+(\s)
  +\vartheta^+\bar\vartheta^+ \p_{--}\p_{++} g(\s)
  ~,
\eeqn
\esubeq
where, thanks to \eqref{02Gppmm}, it holds
\be
\p_{++} G_{--}+\p_{--} G_{++}=0 
~.
\ee


 \section{Generalized Smirnov-Zamolodchikov type composite operators} 
 \label{SZgeneralization}

As already stressed in the main body of the paper, 
one of the important properties of the operators inducing the bosonic $T\bar J$ and $J\bar T$ deformations \cite{Guica:2017lia}
 is to be of Smirnov-Zamolodchikov type \cite{Smirnov:2016lqw}, see $O^{\rm SZ}_{s+s'}(\s)$ defined in equation 
 \eqref{AABB}.
As such, 
despite being composite irrelevant operators, they prove to be free of short distance
singularities and well-defined by a point splitting procedure, as for the analysis in \cite{Smirnov:2016lqw}.
 In the supersymmetric cases that we have studied in this paper, 
 the $T\bar J$ and $J\bar T$ operators
 prove to be supersymmetric descendants of other operators. 
In particular, in this section we will restrict to  the $T\bar J$ and $J\bar T$ operators
that arise as full superspace integrals of some primary operators. In this case, the deformation operators
sit at the bottom of a long supersymmetric multiplet. If supersymmetry is not broken by quantum effects, 
the entire multiplet should be well defined by a point splitting regularization, not only its bottom component.
This is for instance the case for the supersymmetric $T\bar T$ deformations
studied in \cite{Baggio:2018rpv,Chang:2018dge,Jiang:2019hux}. 
Another remarkable feature of the deformation operators is that they are all invariant under improvement transformations of the 
(supersymmetric) currents. As we will see, these are features that hold also for the supersymmetric
$T\bar J$ and $J\bar T$ operators that we have introduced in section \ref{sec:TJbarsusy}.
The way we will show this here, is to actually notice that all  the supersymmetric
$T\bar T$, $T\bar J$ and $J\bar T$ operators belong to a class of
composite operators that generalizes the Smirnov-Zamolodchikov one. 
After describing such a general pattern, we will discuss the well-definedness properties of 
 the supersymmetric primary operators introduced in this paper which we believe extend to the general case
 of the operators defined below by eq.~\eqref{Ogeneral}.

\subsection{Generalized Smirnov-Zamolodchikov operators}
\label{GSZ1}

 It turns out that all    the supersymmetric
$T\bar T$, $T\bar J$ and $J\bar T$ primary operators studied so far in the literature
 fit into the following general pattern:
     \be
\mathcal  O (\z)= \mathcal A(\z)\mathcal  B(\z)- s\mathcal X(\z)\mathcal Y(\z)
~.
\label{Ogeneral}
 \ee
 Here $\mathbb L,\mathbb R$ are superspace differential  operators  
  $\mathbb L,\mathbb R \in \{ \cD_+,\cD_-,\p_\pp,\p_\mm,\p_\pp\cD_+, \cdots \} $ 
  and 
  $\mathcal A,\mathcal B,\mathcal X,\mathcal Y$ are superfields
  satisfying conservation equations of the following type
 \be\label{ABCDLR}
\mathbb L\mathcal  A=  \mathbb R\mathcal   Y
~, \qquad  \mathbb L\mathcal  X=\mathbb R\mathcal  B
~.
 \ee
 This generalizes the Smirnov-Zamolodchikov type of composite operators which corresponds to the case
  $\mathbb L=\p_{--},\mathbb R=\p_{++}$ and $s=1$.
 
 To study some of the properties of these operators,
  we introduce $| A |$ to denote twice of the spin of $A$ 
  which can be either a superfield or a differential operator. 
  Essentially it is given by the sum of $+$ and $-$ indices. For example
 \be
 |\cD_+|=|\cJ_+|=1
 ~, \qquad  |\p_{--} |=|\cG_{--}|=-2
 ~, \qquad
 \cdots
 \qquad
 etc.
 \ee
This satisfies
 \be
 | A  B|=| A|+| B|~, \qquad (-)^{|A|}= (-)^{-| A|}
 ~.
 \ee

 We would first like to understand the behavior of $\mathcal O$ under improvement transformations.%
 \footnote{We refer the reader to \cite{Baggio:2018rpv,Chang:2018dge,Jiang:2019hux} for the improvement 
 transformations of the various stress-tensor multiplets. The flavor current multiplets satisfy similar improvement transformations
 which we have not analyzed in detail in our paper. 
 For our scopes here, it will suffice to use the abstract description given in this appendix.} 
 Suppose $ \mathbb L,  \mathbb R$ are either commuting or anti-commuting~%
 \footnote{It should be noted that \eqref{LRcomm} may not be satisfied, for example in the $\cN=(0,2)$ $J\bar T$ deformation. }
  \be\label{LRcomm}
 \mathbb L \mathbb R =r\mathbb R \mathbb L ~, \qquad r=\pm1
 ~.
 \ee
 Then we can have the following improvement transformations which leave the constraints  \eqref{ABCDLR} invariant:
\bsubeq
  \beqn
\mathcal  A&&\rightarrow \mathcal   A'=\mathcal A+\mathbb R\mathcal U
~, \qquad 
\mathcal Y\rightarrow \mathcal Y'=\mathcal Y+r\mathbb L\mathcal U~,
 \label{ADtsf}
  \\
\mathcal   X&&\rightarrow \mathcal X'=\mathcal X+\mathbb R\mathcal V
~, \qquad 
\mathcal B\rightarrow \mathcal B'=\mathcal B+r\mathbb L\mathcal V ~.
  \label{BCtsf}
 \eeqn
\esubeq
 
An explicit calculation shows that under \eqref{ADtsf}, $\mathcal O$ transforms as
 \be
\mathcal   O \rightarrow\mathcal  O'=\mathcal A' \mathcal B-s\mathcal X\mathcal Y' 
=\mathcal O-(-)^{|\mathbb R|\cdot |\mathcal U|} \mathcal U \Big( \mathbb R \mathcal B- sr t \mathbb L \mathcal X   \Big) 
 +\mathbb L(\cdots)+\mathbb R(\cdots)
 ~,
 \ee
 where 
 \be
 t=(-)^{|\mathbb L|^2-|\mathbb L|\cdot |\mathbb R|+|\mathcal B|\cdot |\mathcal Y|}
 ~.
 \ee
 If   $srt=1$, then using  \eqref{ABCDLR}  gives 
 \be
\mathcal   O'=\mathcal  O+  \mathbb L(\cdots)+\mathbb R(\cdots)   ~,
 \ee
where $ \mathbb L(\cdots),\mathbb R(\cdots) $ are superspace total derivatives and have 
no effect after performing the superspace integral. 
 Then, the deformation operator is invariant 
 under improvement transformation. One can similarly check that $srt=1$ also ensures the 
 improvement invariance under \eqref{BCtsf}.

For the reader's convenience, let us now
list all the supersymmetric primary operators, together with the defining current multiplets
 with $\cN=(0,1)$, $\cN=(1,1)$ and $\cN=(0,2)$ supersymmetry, that we have either  constructed in this paper
or that first appeared in 
the following references \cite{Baggio:2018rpv,Chang:2018dge,Jiang:2019hux}.\footnote{We refer the reader to \cite{Chang:2019kiu}
for $\cN=(2,2)$ $T\bar{T}$ deformations that share similar properties.}
All the following operators are of the form given by eq.~\eqref{Ogeneral}:
  \begin{itemize}
 \item  $(0,1)$ $T\bar T$:
\bsubeq
   \be
 \cO_{ - }^{T\bar T}=  \cT_{----} \cJ_{+++} -\cT \cJ_-
 ~,
 \label{appendixTTbar01}
 \ee
 \beqn
\cD_+ \cT_{----} &=& \ri \p_{--} \cJ_- 
~,\\
\cD_+ \cT &=& \ri \p_{--} \cJ_ {+++} 
~;
 \eeqn
 \esubeq

\item   $(1,1)$ $T\bar T$:
 \bsubeq 
   \be
 \cO ^{T\bar T}=  \cJ_{- --} \cJ_{+++} -\cJ_+ \cJ_-
 \label{appendixTTbar11}
  ~,
 \ee
 \beqn
\cD_+ \cJ_{---} &=&\cD_- \cJ_- 
~,\\
  \cD_+ \cJ_+ &=&  \cD_- \cJ_{+++} 
  ~;
 \eeqn
 \esubeq

\item  $(0,2)$ $T\bar T$:
\bsubeq
 \beqn
 \cO_{-- }^{T\bar T}&=&   \cT_{----} \cS_{++} -\bar \cW_- \cW_- 
\nonumber\\&=& \cT_{----} \cS_{++} + \frac12(\bar \cW_-  - \cW_-   )(\bar \cW_-  - \cW_-   )
 ~,
  \label{appendixTTbar02}
 \eeqn
 \beqn
 (\cD_+- \bar\cD_+) \cT_{----} &=&  \p_{--} \Big( \frac12(\bar \cW_-  - \cW_-   )   \Big) 
 ~,\\
  (\cD_+- \bar\cD_+)(\bar \cW_-  + \cW_-   )&=&  \p_{--}\cS_{++}
  ~;
 \eeqn
  \esubeq

\item  $(0,1)$ $J\bar T$:
\bsubeq
\bea
\cO_+^{J\bar{T} } 
& =&
\cJ_{+++} \cG_{--} 
- \cG_{++}  \cJ_- 
~,
 \label{appendixJTbar01}
\eea
\beqn
\p_{--} \cJ_{+++} &=& -\p_{++} \cJ_- 
~,\\
\p_{--} \cG_ {++} &=& -\p_{++} \cG_{--} 
~;
\eeqn
 \esubeq

\item  $(0,1)$ $T\bar J$:
\bsubeq
 \be
 \cO_{---}^{{T}\bar{J}} 
=
\cT_{----} \cG_{+} 
- \cG_{--}  \cJ_-
~,
 \label{appendixTJbar01}
 \ee
  \beqn
 \cD_+ \cT_{----} &=& \ri \p_{--} \cJ_- 
 ~,\\
  \cD_+  \cG_{--}&=& \ri \p_{--}\cG_+  
  ~;
 \eeqn
  \esubeq

\item  $(1,1)$ $J\bar T$:
\bsubeq
 \be
\cO_{++}^{J\bar{T} } =\cJ_\ppp \cG_-  - \cG_+  \cJ_+ 
~,
 \label{appendixJTbar11}
 \ee
  \beqn
 \cD_- \cJ_{ +++} &=& \cD_+\cJ_+ 
 ~,\\
 \cD_- \cG_{ +} &=& \cD_+\cG_- 
 ~;
 \eeqn
 \esubeq

\item  $(1,1)$ $T\bar J$:
\bsubeq
 \be
\cO_{--}^{T\bar{J} } =\cJ_\mmm \cG_+  - \cG_-  \cJ_-
~,
 \label{appendixTJbar11}
 \ee
  \beqn
 \cD_+ \cJ_{---} &=& \cD_-\cJ_-
 ~,\\
 \cD_+ \cG_{-} &=& \cD_-\cG_+
 ~;
 \eeqn
 \esubeq

\item $(0,2)$ $J\bar T$:~%
\footnote{Remember also that in the case of an $\cR$-multiplet, the $\cN=(0,2)$ $J\bar T$ operator
 is equivalent to
 \be
\cO_{\cR}^{J\bar{T} }(\z)=\cR_{++}(\z)\cG_{--}(\z) -  \cR_{--}(\z) \cG_{++}(\z)
~,
\label{02JTbar-2}
\ee
which is of Smirnov-Zamolodchikov type.
 }

\bsubeq\label{02JTbar0}
 \be\label{02JTbar}
\cO ^{J\bar{T} } = \cG_{--} \cS_{++}- 2  \cT\cG
~,
\ee
 \beqn
 \cD_+ \cG_{--} &=&-\ri \p_{--}\cD_+\cG 
 ~,\\
  \cD_+  (2\cT)&=&-\ri \p_{--}\cD_+\cS_{++} 
  ~;
 \eeqn
  \esubeq

\item $(0,2)$ $T\bar J$ (in term of the $\cR$-multiplet):
\bsubeq  \label{appendixTJbar020}
\be
 \cO_{----}^{T\bar J}=\cT_{----}\cG- \cG_{--} \cdot \frac12 \cR_{--} 
 ~,
  \label{appendixTJbar02}
\ee
 \beqn\label{cons02-222}
  \cD_+ \cT_{----} &=&  -\ri  \p_{--} \cD_+ \Big( \frac{1}{2}\cR_{--}  \Big) 
  ~,\\
   \cD_+ \cG_{--} &=&-\ri \p_{--}\cD_+\cG 
   ~.
 \eeqn  
  \esubeq

\end{itemize}

 \subsection{Well-definedness of   the composite  operators } 
 \label{wellDef}

Of the nine operators listed above, 
we already know that three of them, specifically the operators in eq.~\eqref{appendixTTbar01}, \eqref{appendixTTbar11}
and \eqref{appendixJTbar01} are well-defined (meaning free of short distance singularities in a point-splitting regularization scheme)
since they are of Smirnov-Zamolodchikov type.
Moreover, for the $\cN=(0,2)$ $T\bar{T}$ operator, eq.~\eqref{appendixTTbar02}, we have shown in \cite{Jiang:2019hux}
that  well-definedness can be proven by using supersymmetry and  
point splitting arguments completely analogues of the ones used in \cite{Zamolodchikov:2004ce,Smirnov:2016lqw}.
It turns out that the same arguments apply also to the other operators listed in 
\eqref{appendixTTbar01}--\eqref{appendixTJbar02}  that are not of Smirnov-Zamolodchikov's type. 
For this reason, we will refer the reader to \cite{Zamolodchikov:2004ce,Smirnov:2016lqw} and \cite{Jiang:2019hux}
for details and simply indicate what are the sufficient conditions required to infer well-definedness of the composite operators.
We also believe these arguments might work to prove in general that operators 
of the form \eqref{Ogeneral} satisfying \eqref{ABCDLR} are well-defined.

The heart of the arguments given in \cite{Jiang:2019hux} generalizing  \cite{Zamolodchikov:2004ce,Smirnov:2016lqw} 
was based on the following steps:
\begin{itemize}
\item[{i)}] Define an appropriate bilocal point-splitted version of the composite $\cO(\z)=\cO(\s,\vartheta)$ operator 
whose $\vartheta=0$ 
component, $\cO(\s)=\cO(\z)|_{\vartheta=0}$, defines the supersymmetric primary operator. 
Specifically, for the operators 
of the type
\eqref{Ogeneral}
listed above within eqs.~\eqref{appendixTTbar01}--\eqref{appendixTJbar02}
it suffices to consider
the bilocal superspace operator  given by
\be
\mathcal  O (\z,\z')= \mathcal A(\z)\mathcal  B(\z')- s\mathcal X(\z')\mathcal Y(\z)
~,
\ee
and its $\vartheta=\vartheta'$ limit
\be
\mathcal  O (\s,\s';\vartheta)=\Big{[} \mathcal A(\s,\vartheta)\mathcal  B(\s',\vartheta')
- s\mathcal X(\s',\vartheta')\mathcal Y(\s,\vartheta)\Big{]}|_{\vartheta=\vartheta'}
~.
\ee
Since divergencies cannot occur in the expansions of the Grassmann 
$\vartheta$ and $\vartheta'$ coordinates, the operator $\mathcal  O (\s,\s';\vartheta)$
is the appropriate point-splitted regulated version of the composite superspace operator $\cO(\z)$.

\item[{ii)}] Prove, by using the superspace covariant derivatives algebra, the conservation equations
\eqref{ABCDLR} (and their implications) 
and ``integrations by parts'',
that the bilocal operator satisfies a relation of the following type
\bea
\pa_{\pm\pm}\cO(\z,\z')
&=&
0
+\,\textrm{EoMs}
+(\pa+\pa')[\cdots]
+(\cD+\cD')[\cdots]
~.
\label{pammO-0_pappO-00}
\eea
Here with ``EoMs'' we again refer to terms that are identically zero once the conservation equations
for the current multiplets  are used while with the last two terms
in \eqref{pammO-0_pappO-00} we  indicate terms that are
superspace total derivatives, such as for example 
the vector derivatives $(\pa_{\pm\pm}+\pa'_{\pm\pm})$ 
or, for example, the spinor derivatives $(\cD_++\cD'_+)$, $(\cD_-+\cD'_-)$, $etc$, acting on bilocal operators.

\item[{iii)}]
When we consider the coincident limit  $\vartheta=\vartheta'$ in the Grassmann coordinates,
equation \eqref{pammO-0_pappO-00} implies
\bea
\pa_{\pm\pm}\cO(\s,\s';\vartheta)
&=&
0
\,+\,\textrm{EoMs}
\,+\,[P,\cdots]
\,+\,[ Q,\cdots]
~,
\label{paO}
\eea
where $[P,\cdots]$ and $[ Q,\cdots]$ schematically indicate a translation and supersymmetry transformation of some bilocal 
superfield operator.
Assuming that the model under consideration has preserved translation invariance and supersymmetry,
by using an extension of the OPE arguments of \cite{Zamolodchikov:2004ce,Smirnov:2016lqw},
 one can show that eq.~\eqref{paO} implies \cite{Jiang:2019hux}
\be
\cO(\s,\s';\q)
=
\cO(\z)
\,+\,
{\rm derivative~terms}
~.
\ee 
Here ``derivative terms'' indicate superspace covariant derivatives 
acting on local superfield operators while $\cO(\z)$ arises from the regular, non-derivative part of the OPE of $\cO(\s,\s';\vartheta)$. 
For this reason, up to total derivatives which
for instance do not contribute when the operator is integrated over the full superspace, 
 $\cO(\s,\s';\q)$ is free of short distance singularities  in $\s\to\s'$.
This concludes the arguments of well-definedness of 
\cite{Jiang:2019hux,Zamolodchikov:2004ce,Smirnov:2016lqw}.

\end{itemize}

Let us give an example of the calculation that leads to eq.~\eqref{pammO-0_pappO-00}. 
The simplest case is the
$\cN=(0,1)$ $T\bar J$
for which we define the bilocal operator
 \be
 \cO_{---}^{{T}\bar{J}} (\z,\z')
=
\cT_{----}(\z) \cG_{+}(\z')
- \cG_{--}(\z')  \cJ_-(\z)
~.
\ee
We compute
 \bea
\pa_{++} \cO_{---}^{{T}\bar{J}} (\z,\z')
&=&
\pa_{++}\cT_{----}(\z) \cG_{+}(\z')
+\cJ_-(\z) \pa'_{++}\cG_{--}(\z')
\non\\
&&
-(\pa_{++}+\pa'_{++})\big( \cJ_-(\z) \cG_{--}(\z') \big)
\non\\
&=&
\ri\cD_+\cD_+\cT_{----}(\z) \cG_{+}(\z')
+ \ri\cJ_-(\z)\cD'_+\cD'_+\cG_{--}(\z')
\non\\
&&
-(\pa_{++}+\pa'_{++})\big( \cJ_-(\z) \cG_{--}(\z') \big)
\non\\
&=&
-\cD_+\pa_{--}\cJ_{-}(\z) \cG_{+}(\z')
-\cJ_-(\z)\cD'_+\pa'_{--}\cG_+(\z')
\non\\
&&
+\ri\cD_+\big(\cD_+\cT_{----}(\z)-\ri\pa_{--}\cJ_-(\z)\big) \cG_{+}(\z')
\non\\
&&
+ \ri\cJ_-(\z)\cD'_+\big(\cD'_+\cG_{--}(\z')-\ri\pa_{--}\cG_+(\z')\big)
\non\\
&&
-(\pa_{++}+\pa'_{++})\big( \cJ_-(\z) \cG_{--}(\z') \big)
~,
\eea
where we used  $\pa_{++}=\ri\cD_+\cD_+$, made some ``integration by parts'', and completed terms that are zero 
once the conservation equations for the current multiplets are used.
If we ``integrate by parts'' both the $\cD_+$ and $\pa_{--}$ derivatives in the first line of the last equivalence we obtain
\bea
\pa_{++} \cO_{---}^{{T}\bar{J}} (\z,\z')
&=&
\cJ_{-}(\z) \pa'_{--}\cD'_+\cG_{+}(\z')
-\cJ_-(\z)\pa'_{--}\cD'_+\cG_+(\z')
\non\\
&&
+\ri\cD_+\big(\cD_+\cT_{----}(\z)-\ri\pa_{--}\cJ_-(\z)\big) \cG_{+}(\z')
\non\\
&&
+ \ri\cJ_-(\z)\cD'_+\big(\cD'_+\cG_{--}(\z')-\ri\pa_{--}\cG_+(\z')\big)
\non\\
&&
-(\pa_{++}+\pa'_{++})\big( \cJ_-(\z) \cG_{--}(\z') \big)
\non\\
&&
-(\pa_{--}+\pa'_{--})\big(\cD_+\cJ_{-}(\z) \cG_{+}(\z')\big)
\non\\
&&
+(\cD_++\cD'_+)\big(\cJ_{-}(\z) \pa'_{--}\cG_{+}(\z')\big)
~,
\eea
where the first term is identically zero, the second and third line are zero once used the conservation equations,
while the last three lines are total derivatives. 
A very similar calculation shows that the following result holds
 \bea
\pa_{--} \cO_{---}^{{T}\bar{J}} (\z,\z')
&=&
-\ri\cT_{----}(\z)\big(\cD'_+\cG_{--}(\z')- \ri\pa'_{--}\cG_{+}(\z')\big)
\non\\
&&
-\ri \big(\cD_+\cT_{----}(\z)-\ri\pa_{--}\cJ_-(\z)\big)  \cG_{--}(\z')
\non\\
&&
+(\pa_{--}+\pa'_{--})\big(\cT_{----}(\z) \cG_{+}(\z')\big)
\non\\
&&
+\ri(\cD_++ \cD'_+)\big(\cT_{----}(\z)  \cG_{--}(\z')\big)
~,
\eea
which, again, is zero up to total derivatives and terms that cancel once the conservation equations are used.
These show that the composite bilocal operator $\cO_{---}^{{T}\bar{J}} (\z,\z')$ satisfies 
eq.~\eqref{pammO-0_pappO-00}
\be
\pa_{\pm\pm}\cO_{---}^{{T}\bar{J}} (\z,\z')
=
0
+\,\textrm{EoMs}
+(\pa+\pa')[\cdots]
+(\cD+\cD')[\cdots]
~,
\ee
and then $\cO_{---}^{{T}\bar{J}} (\z)$ is well-defined.

Similar calculations hold for the operators defined by the equations
\eqref{appendixJTbar11} and \eqref{appendixTJbar11} in the $\cN=(1,1)$  case, 
while the $\cN=(0,2)$ $J\bar T$ operator of
eq.~\eqref{02JTbar}, in the case of an $\cR$-multiplet
 does not need any significant analysis since it is equivalent to a Smirnov-Zamolodchikov type operator,
see equation \eqref{02JTbar-2}
(the same is true for the bilocal forms of the $\cN=(0,2)$ $J\bar T$ operators).
We leave as an exercise to the reader to prove that \eqref{pammO-0_pappO-00} holds for 
\eqref{appendixJTbar11} and \eqref{appendixTJbar11}.

We are left with the $\cN=(0,2)$ $T\bar J$ operator, eq.~\eqref{appendixTJbar02}, which assume the existance of an $\cR$-multiplet,
and the $\cN=(0,2)$ $J\bar T$ operator of eq.~\eqref{02JTbar} for a general $\cS$-multiplet.
Let's focus on the $T\bar J$ case, the general $\cN=(0,2)$ $J\bar T$ analysis goes along the same lines.
By doing some straightforward manipulations similar  to the ones used above one can prove  the following relation
\bea
\pa_{++} \cO_{----}^{T\bar J}(\z,\z')
&=&
-\frac{\ri}{2}\cD_+\Big(\cDB_+\cT_{----}(\z)-\frac{\ri}{2}\cDB_+\pa_{--}\cR_{--}(\z)\Big)\cG(\z')
\non\\
&&
-\frac{\ri}{2}\cDB_+\Big(\cD_+\cT_{----}(\z)+\frac{\ri}{2}\cD_+\pa_{--}\cR_{--}(\z)\Big)\cG(\z')
\non\\
&&
-\frac{\ri}{4}\cR_{--}(\z)\cD'_+\Big(\cDB'_+ \cG_{--}(\z')-\ri\cDB'_+\pa'_{--}\cG(\z')\Big)
\non\\
&&
-\frac{\ri}{4}\cR_{--}(\z)\cDB'_+\Big(\cD'_+ \cG_{--}(\z')+\ri\cD'_+\pa'_{--}\cG(\z')\Big)
\non\\
&&
-\hf(\pa_{++}+\pa'_{++}) \cR_{--}(\z) \cG_{--}(\z')
\non\\
&&
-\frac{1}{4}(\pa_{--}+\pa'_{--})\Big(\cD_+\cR_{--}\cDB'_+\cG(\z')-\cDB_+\cR_{--}\cD'_+\cG(\z')\Big)
\non\\
&&
+\frac{1}{4}(\cD_++\cD'_+)\Big(
\pa_{--}\cDB_+\cR_{--}(\z)\cG(\z')
+\cR_{--}(\z)\pa'_{--}\cDB'_+\cG(\z')
\Big)
\non\\
&&
-\frac{1}{4}(\cDB_++\cDB'_+)\Big(
\cR_{--}(\z)\pa'_{--}\cD'_+\cG(\z')
+\pa_{--}\cD_+\cR_{--}(\z)\cG(\z')\Big)
~,~~~~~~~~~
\eea
which, as expected, is of the form
\be
\pa_{++}\cO_{----}^{{T}\bar{J}} (\z,\z')
=
0
+\,\textrm{EoMs}
+(\pa+\pa')[\cdots]
+(\cD+\cD')[\cdots]
~.
\ee
The analysis of $\pa_{--} \cO_{----}^{T\bar J}(\z,\z')$ is more intricate since it is clear that $\pa_{--} $ acting on any
superfields in the current multiplets can not be directly simplified by using the conservation equations
\eqref{cons02-222}.
As a way around, we assume that the anti-chirality constraints (and their complex conjugates)
in \eqref{cons02-222}
can be solved on-shell in terms of two local composite complex superfields $\cP_{-----}(\z)$ and $\cP_{---}(\z)$ as
\bsubeq
 \bea
\Big( \cT_{----} +\frac{\ri}{2}  \p_{--} \cR_{--} \Big)&=&\cD_+\cP_{-----}
  ~,\\
   \Big(\cG_{--}+\ri \p_{--}\cG \Big)&=&\cD_+\cP_{---}
      ~,
 \eea
  \esubeq
which imply
\bsubeq
 \bea
\cT_{----} &=&\hf\big(\cD_+\cP_{-----}+\cDB_+\bar{\cP}_{-----}\big)
  ~,\\
 \p_{--} \cR_{--}&=&-\ri\big(\cD_+\cP_{-----}-\cDB_+\bar{\cP}_{-----}\big)
  ~,\\
\cG_{--}&=&\hf\Big(\cD_+\cP_{---}+\cDB_+\bar{\cP}_{---}\Big)
      ~,
\\
\p_{--}\cG &=&-\frac{\ri}{2}\big(\cD_+\cP_{---}-\cDB_+\bar{\cP}_{---}\big)
~,
    \eea
  \esubeq
where $\bar{\cP}_{-----}=\overline{(\cP_{-----})}$
and $\bar{\cP}_{---}=\overline{(\cP_{---})}$.
By using the decomposition in terms of the prepotential superfields 
$\cP_{-----}(\z)$ and $\cP_{---}(\z)$ we can analyse $\pa_{--} \cO_{----}^{T\bar J}(\z,\z')$. A straightforward calculation 
similar to the previous cases
shows that it holds
\bea
\pa_{--} \cO_{----}^{T\bar J}(\z,\z')
&=&
(\pa_{--}+\pa'_{--})\big( \cT_{----}(\z)\cG(\z')\big)
\non\\
&&
+\frac{\ri}{2}(\cD_++\cD'_+)\big(\cP_{-----}(\z)\cD'_+\cP_{---}(\z')\big)
\non\\
&&
-\frac{\ri}{2}(\cDB_++\cDB'_+)\big(\bar{\cP}_{-----}(\z)\cDB'_+\bar{\cP}_{---}(\z')\big)
~,
\eea
which is an equation of the form
\be
\pa_{--}\cO_{----}^{{T}\bar{J}} (\z,\z')
=
0
+\,\textrm{EoMs}
+(\pa+\pa')[\cdots]
+(\cD+\cD')[\cdots]
~,
\ee
as expected. This finalizes the analysis of the well-definedness for the $\cN=(0,2)$ ${{T}\bar{J}}$ operator.
The reader can use the same on-shell resolution of the chirality constraints to show that the same analysis can be performed
with the $\cN=(0,2)$ $J\bar T$ operator of eq.~\eqref{02JTbar} for a general $\cS$-multiplet.
In fact, the arguments are almost identical considering the same structures of \eqref{02JTbar0} and \eqref{appendixTJbar020}.

To conclude this section we stress, once more, that despite we have not yet attempted to prove that
the generalized Smirnov-Zamolodchikov operators defined in eq.~\eqref{Ogeneral} are well-defined in general, we expect
that a proof will develop along the lines of the cases analyzed so far.


\bibliographystyle{JHEP}
\bibliography{JTbar__refs}

\end{document}